\documentclass{elsart}
\def\half{{\textstyle{1\over2}}}
\def\2{{1\over 2}}

\newcommand{\rf}[1]{(\ref{#1})}

\renewcommand{\t}{\tilde}
\renewcommand{\b}{\bar}
\newcommand{\p}{\partial}
\newcommand{\bp}{\bar{\partial}}
\newcommand{\tc}{\tilde c }

\usepackage{amssymb}
\journal{Nuclear Physics B}
\begin{document}
\begin{frontmatter}
\title{BRST, Generalized Maurer-Cartan Equations and CFT}
\author{Anton M. Zeitlin}
\author{}
\ead[url]{http://pantheon.yale.edu/\~{  }az84, http://www.ipme.ru/zam.html}
\address{Department of Mathematics,\\
Yale University,\\
442 Dunham Lab, 10 Hillhouse Ave\\
New Haven, CT 06511\\
\ead {anton.zeitlin@yale.edu, zam@math.ipme.ru}
\vspace{5mm}
St. Petersburg Department of \\
Steklov Mathematical
Institute,\\
 Fontanka, 27, \\
 St. Petersburg, 191023, Russia}
\begin{abstract}
The paper is devoted to the study of BRST charge in perturbed two dimensional conformal field theory. 
The main goal is to write the operator equation expressing 
the conservation law of BRST charge in perturbed theory in terms of purely algebraic operations on the 
corresponding operator algebra, which are defined via the OPE. 
The corresponding equations are constructed and their symmetries are studied up to the second order in 
formal coupling constant. It appears that the obtained equations can be interpreted 
as generalized Maurer-Cartan ones. 
We study two concrete examples in detail: the bosonic nonlinear sigma model and perturbed first order theory. 
In particular, we show that the Einstein equations, 
which are the conformal invariance conditions for both these perturbed theories, 
expanded up to the second order, can be rewritten in such generalized Maurer-Cartan form. 
\end{abstract}
\end{frontmatter}

\section{Introduction}
The study of perturbed Conformal Field Theories and String Theory in nontrivial backgrounds is a very 
complicated task to which many papers were addressed during more than twenty years.
One of the most important questions, which emerges in the study of nontrivial backgrounds for String Theory, 
is: whether this background (e.g. nontrivial metric, dilaton, etc) gives a conformal invariant quantum field 
theory. The answer to this question was given in 80-s by means of calculation of 
the $\beta$-function of the associated nonlinear sigma-model. 
For the bosonic sigma models with the action  
\begin{eqnarray}
S=\frac{1}{4\pi\alpha'}\int d^2 z (G_{\mu\nu}(X)+B_{\mu\nu}(X))\p X^{\mu}\bp X^{\nu},
\end{eqnarray}
considered on a flat worldsheet, the conditions of conformal 
invariance at zero order in $\alpha'$ appeared to be the Einstein equations (see e.g. \cite{eeq}-\cite{hull}):
 \begin{eqnarray}
&&\label{e1}R_{\mu\nu}={1\over 4} H_{\mu\lambda\rho}H_{\nu}^{\lambda\rho}-2\nabla_{\mu}
\nabla_{\nu}\Phi,\\
&&\label{e2}
\nabla_{\mu}H^{\mu\nu\rho}-2(\nabla_{\lambda}\Phi)H^{\lambda\nu\rho}=0,
\end{eqnarray}
where $\Phi$ is a dilaton field. 
Usually these equations are accompanied with another ``dilatonic'' equation 
\begin{eqnarray}
\label{e3}
4(\nabla_{\mu}\Phi)^2-4\nabla_{\mu}\nabla^{\mu}\Phi-
R+{1\over 12} H_{\mu\nu\rho}H^{\mu\nu\rho}=const,
\end{eqnarray}
which is the consequence of \rf{e1}, such that constant in the r.h.s. of \rf{e3} is identified  
with central charge of the corresponding CFT \cite{hull}.  
However, this approach gave only superficial 
relations with traditional operator approach of String theory. 
The question is how these conditions of conformal 
invariance and their (target-space) symmetries could be reformulated in this natural operator approach.\\
\hspace*{5mm}
In this paper, we try to make first steps in this direction. 
As a key tool we consider the accurately defined 
deformed BRST operator in the perturbed theory. For both of our examples,  
the first order \cite{kap}-\cite{nekrasov} and the usual second order sigma models, 
the equation of BRST charge conservation,  
up to the second order in the (formal) coupling constant $t$, leads to the Einstein equations expanded 
up to the second order in this formal parameter. 
We also show that these operator equations can be interpreted as generalized Maurer-Cartan ones, and  
find the operator formulation of associated symmetries.\\ 
\hspace*{5mm}
The outline of the paper is as follows. 
In section 2, motivated by the simple analogy with Quantum Mechanics 
we postulate the equation expressing the (regularized) conservation law of current in the background of the perturbation 
2-form $\phi^{(2)}=\sum^{\infty}_{n=1}t^n\phi_n^{(2)}$ expanded with respect to the formal coupling constant $t$: 
\begin{eqnarray}\label{mint}
&&[Q,\phi_1^{(2)}](z)=d\psi_1^{(1)}(z),\nonumber\\ 
&&[Q,\phi_2^{(2)}](z)+\frac{1}{2\pi i}\int_{C_{\epsilon, z}}\psi^{(1)}_1\phi_1^{(2)}(z)=d\psi_2^{(1)}(z), 
\end{eqnarray}
where $C_{\epsilon, z}$ is a small circle of radius $\epsilon$ around $z$ and 
$\psi^{(1)}=\sum^{\infty}_{n=1}t^n\psi_n^{(1)}$ is a deformation of a 1-form current $J$, such that 
$J$ is conserved in the non perturbed theory; the charge $Q$ in the equations 
above is a conserved charge associated in the usual way with current $J$.\\
\hspace*{5mm}
Since we have an $\epsilon$-dependent operation 
in the second equation from \rf{mint} it is reasonable to make $\phi_2^{(2)}$ and $\psi_2^{(1)}$ also $\epsilon$-dependent 
(the renormalization). \\
\hspace*{5mm}
Then we reformulate these two equations in the Maurer-Cartan-like form:
\begin{eqnarray}
&&D\Phi_1(z,\theta)=0,\nonumber\\
&&D\Phi_2(z,\theta)+\int_{C_{\epsilon, z}}\Phi_1(w,\theta)\Phi_1(z,\theta)=0,
\end{eqnarray}
where $D=d+\theta Q$, $\Phi_i=\phi^{(2)}_i+\theta  \psi^{(1)}_i$, $\theta$ is some Grassmann number 
anticommuting with $d$, and $\int_{C_{\epsilon, z}}$ is equal to zero when it is applied to the two form $\phi^{(2)}_1$. 
We also discuss the symmetries of these equations, which are of the form:
\begin{eqnarray}
&&\delta\Phi_1=-D\Lambda_1,\nonumber\\
&&\delta\Phi_2=-D\Lambda_2+M_{\epsilon}(\Lambda_1,\Phi_1), 
\end{eqnarray}
where $\Lambda_i=\xi_i^{(1)}+\theta \alpha_i^{(0)}$ and $M_{\epsilon}$ is some bilinear operation properly defined.\\ 
\hspace*{5mm}In section 3, we specify the type of current/charge, that we deform, and the type of the perturbation to 
get more close to 
the sigma models (see sections 4,5). The deforming charge we take is a BRST operator, 
which can be constructed for any CFT and we put some conditions on the OPEs of perturbing operator.  
We choose some ansatz for $\psi^{(1)}$, motivated by the properties of the BRST operator and 
dimensional reasons. 
Deriving the concrete equations 
relating various operator products, we find some ambiguity in the solutions of \rf{mint}. 
The 1-form $\psi_1^{(1)}$ is defined up to the total derivative (since it enters the equations 
with either a total derivative or with integration over the closed contour) and it is usual for the current, 
but studying the first equation from \rf{mint} we find that 
the solution for $\psi_1^{(1)}$ is defined up to the closed 1-form, which together with constraints 
given by our ansatz has the following explicit form: 
\begin{eqnarray}\label{zterm}
cZ^h_1dz-\t c Z^a_1d\b z,  
\end{eqnarray}
where $Z^h_1$, $Z^a_1$ are correspondingly holomorphic and antiholomorphic operators, 
$c, \t c$ are usual ghost fields entering the BRST operator. This can lead to the ambiguity in the second equation of \rf{mint} since it depends on $\psi_1^{(1)}$. 
However, there is a way to avoid the appearance of such $Z_1$-terms. We put the following constraint 
\begin{eqnarray}
\psi_1^{(1)}=\phi_1^{(1)}\equiv dz[b_{-1},\phi_1^{(0)}]+d\b z[\t b_{-1},\phi_1^{(0)}] 
\end{eqnarray}
for some $\phi_1^{(0)}$ of ghost number 2. This will automatically lead to the absence of 
the terms \rf{zterm}. 
Then, motivated by the properties of BRST operator, one may reexpress 
the equations \rf{mint} in terms of $\phi_i^{(0)}$-operators:
\begin{eqnarray}\label{Mint}
&&[b_{-1},[\t b_{-1},[Q_B,\phi_1^{(0)}]]]=0,\nonumber\\
&&[b_{-1},[\t b_{-1},[Q_B,\phi_2^{(0)}]+\frac{1}{4\pi i}\int_{C_{\epsilon,z}} \phi_1^{(1)}(w)\phi_1^{(0)}(z)]]=0
\end{eqnarray}
where $b_{-1},\t b_{-1}$ are the modes of $b, \t b$-ghost fields entering the BRST-operator and 
$\phi_2^{(0)}$ is some operator of ghost number 2 such that 
$\phi_2^{(2)}$=$dz\wedge d\b z[b_{-1},[\t b_{-1},\phi_2^{(0)}]]$. \\
\hspace*{5mm}
In the following, we will refer to the system \rf{Mint} as {\it Master Equation} since we suppose that 
this system can be unified into one equation:
\begin{eqnarray}
[b_{-1},[\t b_{-1},[Q_B,\phi^{(0)}]+\frac{1}{4\pi i}\int_{C_{\epsilon,z}} \phi^{(1)}(w)\phi^{(0)}(z)+...]]=0
\end{eqnarray}
where dots mean the terms of higher order in $\phi^{(0)}$ and its descendants 
$\phi^{(1)}$ and $\phi^{(2)}$, such that $\phi^{(0)}=t\phi_1^{(0)}+t^2\phi_2^{(0)}+O(t^3)$. 
In the next section, we apply the obtained results to the concrete examples.\\
\hspace*{5mm} 
Section 4 is devoted to study of Master Equation in the case of first order field theory
with the action 
\begin{eqnarray}
S_0=\frac{1}{2\pi\alpha'}\int d^2 z (p_i\bar{\partial} X^{i}+
p_{\bar{i}}{\partial} X^{\bar{i}})
\end{eqnarray}
perturbed by the 2-form 
\begin{eqnarray}
\phi^{(2)}_g=dz\wedge d\b z{\alpha'}^{-1}g^{i\b j}(X,\b X)p_ip_{\b j},
\end{eqnarray}
where $i,\b i=1,...,D/2$ ($D$ is even); $p_i$ ($p_{\b i}$) are (1,0) ((0,1)) primary fields and  $X^i$, $X^{\b i}$ 
are also primary of zero conformal weights.  
This theory (with and without the perturbation above) has many interesting properties and applications (see e.g. 
\cite{witten}-\cite{nekrasov}).\\ 
\hspace*{5mm}We find that in this case the Master Equation leads to the equations \rf{e1} and \rf{e2}, 
where metric and B-field are expressed as follows: 
\begin{eqnarray}
G_{i\bar{k}}=g_{i\bar{k}}, \quad B_{i\bar{k}}=-g_{i\bar{k}},
\end{eqnarray}
expanded up to the second order in the formal parameter. 
This is what we expected, since after (functional) integration over $p_i, p_{\b i}$-variables 
we arrive to the usual sigma model with metric and the B-field given as above. 
In this case Einstein equations are quadratic in the $g^{i\b j}$ tensor field, so our 
second order approximation is exact. 
In the end of this section we study the unexpected (target-space) algebraic structure of these equations. \\
\hspace*{5mm} 
In section 5, the usual sigma model with the action
\begin{eqnarray}
S_{G}=\frac{1}{4\pi\alpha'}\int d^2 z G_{\mu\nu}(X)\p X^{\mu}\bp X^{\nu},
\end{eqnarray}
where $\mu,\nu=1,...,D$ ($D$ is a dimension of the target space), is discussed. 
Expanding $G_{\mu\nu}=\eta_{\mu\nu}-th_{\mu\nu}(X)-t^2s_{\mu\nu}(X)+O(t^3)$ we find that 
the perturbation operators $\phi_{G,i}^{(2)}$ naively have the following expressions:
\begin{eqnarray}
&&\phi_{G,1}^{(2)}=dz\wedge d\b z(2\alpha')^{-1}h_{\mu\nu}(X)\p X^{\mu}\bp X^{\nu},\\
&&\phi_{G,2}^{(2)}=dz\wedge d\b z(2\alpha')^{-1}s_{\mu\nu}(X)\p X^{\mu}\bp X^{\nu}.\nonumber
\end{eqnarray} 
However, analyzing the symmetries and the Master Equation itself, we find that the 
proper expression for $\phi_{G,2}^{(2)}$ should include the {\it bivertex} operator:
\begin{eqnarray}
&&\phi_{G,2}^{(2)}=dz\wedge d\b z(2\alpha')^{-1}(s_{\mu\nu}(X)\p X^{\mu}\bp X^{\nu}+\\
&&\frac{1}{2}h_{\mu\rho}(X)\eta^{\rho\sigma}h_{\nu\sigma}(X)\p X^{\mu}\bp X^{\nu}).\nonumber
\end{eqnarray}
The Master Equation applied to the $\phi_{G,i}^{(2)}$ leads to the equation \rf{e1} (with $B_{\mu\nu}=0$)  
expanded up to the second order in $t$. \\
\hspace*{5mm} 
Section 6 outlines the possible way of construction of the complete Master equation and other directions of 
further study. 
\section{Basic Equations and their Symmetries.}
\subsection{Motivation: Deformed currents and charges}
\noindent Let's consider the charge $Q$ in the euclidean quantum mechanics with the 
hamiltonian $H_0$. If it is conserved, it commutes 
with the hamiltonian: $[Q,H_0]=0$. Let's suppose  
there is a perturbation $H_0\to H=H_0+V$, where $V$ is expanded with respect to 
some formal coupling constant $g$: $V=\sum^{\infty}_{n=1}V_ng^n$. 
The charge $Q$ usually does not commute with $H$ and therefore it 
is not conserved in the perturbed theory. However, let's try to deform the charge $Q\to Q^{def}=Q+\psi$ ($\psi=
\sum^{\infty}_{n=1}\psi_ng^n$) in such 
a way that $Q^{def}$ is conserved, i.e. $[Q^{def}, H]=0$. This leads to the following relation between 
$\psi$ and $V$ at the second order in $g$:
\begin{eqnarray} 
&&[Q,V_1](\tau)=\frac{d}{d\tau}\psi_1(\tau),\nonumber\\
&&[Q,V_2](\tau)+[\psi_1,V_1](\tau)=\frac{d}{d\tau}\psi_2(\tau),
\end{eqnarray}
where we have included time dependence with respect to the hamiltonian $H_0$ (recall that in euclidean Quantum Mechanics, 
the evolution is given as follows $A(\tau)=e^{H_0\tau}Ae^{-H_0\tau}$). 
However, to be well defined, the commutator between $\psi$, $V$ should be regularized. 
One of possible ways to 
do this is to include a compact operator of type $e^{-\epsilon H_0}$ between operators in the commutator, 
i.e. it is reasonable to consider the following pair of equations with shifted time variables inside the commutator:
\begin{eqnarray} \label{qmme}
&&[Q,V_1](\tau)=\frac{d}{d\tau}\psi_1(\tau),\nonumber\\
&&[Q,V_2](\tau)+\psi_1(\tau+\epsilon)V_1(\tau)-V_1(\tau)\psi_1(\tau-\epsilon)=\frac{d}{d\tau}\psi_2(\tau).
\end{eqnarray}
Now let's consider the 2d CFT on the complex plane and a 
conserved current $J=j(z)dz-\t j(z) d\b z$. This means that the following relation 
holds under the correlator
\begin{equation} 
\int_{C} J(y) =0
\end{equation} 
for any closed contour $C$, with the condition that no other operators are inserted inside $C$.
The commutator of the associated conserved charge with any operator $v(z)$ can be  
defined in the following way:
\begin{equation}
[Q,v(z)]=\frac{1}{2\pi i}\int_{C_{r,z}} J(y) v(z),
\end{equation}
where $C_{r,z}$ is a circle contour of radius $r$ around $z$ ($r$ is supposed to be small enough such that 
no other operators are inserted inside $C_{r,z}$). 
Suppose we have perturbed our CFT by a 2-form $\phi^{(2)}=Vdz\wedge d\b z$ such that it is expanded 
with respect to the formal coupling constant $t$
$\phi^{(2)}=\sum^{\infty}_{n=1}t^n\phi_n^{(2)}$. Generically the current $J$ is no longer conserved in 
such a background. So, we need to build something similar to the relations \rf{qmme} in the field theory case.
We propose the following equations providing the conservation of the deformed current 
$J^{def}=J+\psi^{(1)}$ ($\psi^{(1)}=\sum^{\infty}_{n=1}t^n\psi_n^{(1)}$)
in the background of a perturbation $\phi^{(2)}$ up to the second order in $t$:
\begin{eqnarray}\label{me}
&&\label{me1}[Q,\phi_1^{(2)}](z)=d\psi_1^{(1)}(z),\\ 
&&\label{me2}[Q,\phi_2^{(2)}](z)+\frac{1}{2\pi i}\int_{C_{\epsilon, z}}
\psi^{(1)}_1(w)\phi_1^{(2)}(z)=d\psi_2^{(1)}(z),
\end{eqnarray}
where $\epsilon$ is some finite regularization parameter \footnote{The equations \rf{me1}-\rf{me2} also can be 
substantiated by consideration of the deformed current in the background of the perturbation 
series regularized by the point-splitting with splitting parameter $\epsilon$. This problem will be studied 
elsewhere.}.
\subsection{Hidden Maurer-Cartan Structures}
\noindent Before we proceed to the study the equations \rf{me1} and \rf{me2}, we need to define the 
structure of $\phi_i^{(2)}$, $\psi_i^{(1)}$ properly. We make two assumptions, motivated by our basic examples.\\
{\bf Assumption 1.} Any two operators $V,W$ are supposed to have the OPE of such a type:   
\begin{eqnarray}\label{ope}
&&V(z)W(z')=\nonumber\\
&&\sum_{r=-\infty}^{m}\sum_{s=-\infty}^{n}\sum^{\infty}_{i=0}(V,W)^{(r,s)}_i(z')(z-z')^{-r}(\b z-\b z')^{-s}
(\alpha'\ln|(z-z')/\mu|)^i
\end{eqnarray} 
for some $m,n\in \mathbb{Z}$, 
where $\alpha'$ and $\mu$ are some formal parameters, $r,s,i\in \mathbb{Z}$; if 
the operators $V,W$ don't depend on on $\alpha'$ and $\mu$, then 
$(V,W)^{(r,s)}_i$ polynomially depend on $\alpha'$ and don't depend on $\mu$.\\   
{\bf Assumption 2.} The regularization parameter $\epsilon$ is supposed to be small enough in order to the operation in 
\rf{me2} could be calculated via the OPE. We consider $\phi_1^{(2)}$ as an $\epsilon$-independent operator 
and we assume that $\phi_2^{(2)}$ and $\psi^{(1)}_2$ 
can depend on $\epsilon$ and $\ln(\epsilon/\mu)$ in the following way:\\
$\phi^{(2)}_2=\sum^{\infty}_{k=0}\sum^{\infty}_{l=-m}\phi_{2;k,l}^{(2)}\epsilon^l(\ln (\epsilon/\mu))^k$ and 
$\psi_2^{(1)}=\sum^{\infty}_{k=0}\sum^{\infty}_{l=-m}\psi^{(1)}_{2;k,l}\epsilon^l(\ln (\epsilon/\mu))^k$ 
for some finite $m$.\\
\hspace*{5mm}The reason for $\epsilon$-independence of $\phi_1^{(2)}, \psi_1^{(2)}$ is quite obvious, 
since there is no need in $\epsilon$ renormalization at the first order in the coupling constant, while at the 
second order we have $\epsilon$-dependent operation and there inclusion of $\epsilon$-terms is necessary. \\
\hspace*{5mm}After these preparations we switch to the algebraic analysis of the equations \rf{me1} and \rf{me2}. 
Let's define the differential 
\begin{eqnarray}
D\equiv d+\theta Q,
\end{eqnarray}
where $Q$ is a charge associated with some conserved current $J$, $d$ is a de Rham differential, and $\theta$ is a 
Grassmann variable anticommuting with d. Then the equation \rf{me1} has the following form:
\begin{eqnarray}
\label{meu1}
D\Phi_1=0,
\end{eqnarray}
where $\Phi_1\equiv \phi_1^{(2)}+\theta \psi_1^{(1)}$.\\
To describe \rf{me2} in a similar way, we introduce an operation $K_{\epsilon}$ on $\Phi=\phi^{(2)}+\theta \psi^{(1)}$ and 
$\Phi'=\phi'^{(2)}+\theta \psi'^{(1)}$ as follows: 
\begin{eqnarray}
K_{\epsilon}(\Phi,\Phi')(z,\theta)\equiv 1/2(\int_{C_{\epsilon, z}}\Phi(w,\theta)\Phi'(z,\theta)+ 
\int_{C_{\epsilon, z}}\Phi'(w,\theta) \Phi(z,\theta)),
\end{eqnarray}
where integral over $C_{\epsilon, z}$ is equal to zero, acting on 2-forms $\phi^{(2)}$ and $\phi'^{(2)}$.
In other words, we have:
\begin{eqnarray}
&&K_{\epsilon}(\Phi,\Phi')(z, \theta)=\nonumber\\
&&\frac{1}{2}\theta(\frac{1}{2\pi i}\int_{C_{\epsilon, z}}\psi^{(1)}_1(w){\phi'_1}^{(2)}(z)+
\frac{1}{2\pi i}\int_{C_{\epsilon, z}}{\psi'}^{(1)}_1(w)\phi_1^{(2)}(z)).
\end{eqnarray}
Then the equation \rf{me2} allows the following representation:
\begin{eqnarray} 
\label{meu2}
D\Phi_2+K_{\epsilon}(\Phi_1,\Phi_1)=0,
\end{eqnarray} 
where $\Phi_2\equiv \phi_2^{(2)}+\theta \psi_2^{(1)}$.\\
Thus the generalized equation describing the conservation of current should be of the following form:
\begin{eqnarray}\label{genmc} 
D\Phi+K_{\epsilon}(\Phi,\Phi)+...=0,
\end{eqnarray} 
where $\Phi=\sum^{\infty}_{n=1}t^n\Phi_n$ and dots mean the higher operations.
\subsection{Symmetries} 
\noindent The equation \rf{meu1} has the obvious symmetry:
\begin{eqnarray}
\delta\Phi_1=-D\Lambda_1,
\end{eqnarray}
where $\Lambda_1=\theta\alpha_1^{(0)}+\xi_1^{(1)}$ is infinitesimal. 
In components, this looks as follows:
\begin{eqnarray}
\delta\phi_1^{(2)}(z)=-d\xi_1^{(1)}(z), \quad \delta \psi_1^{(1)}(z)=d\alpha^{(0)}(z)-[Q,\xi_1^{(1)}](z).
\end{eqnarray}
The transformations 
\begin{eqnarray}
&&\delta\Phi_1=-D\Lambda_1,\nonumber\\
&&\delta\Phi_2=-D\Lambda_2+M_{\epsilon}(\Lambda_1,\Phi_1), 
\end{eqnarray}
where $\Lambda_2=\theta\alpha_2^{(0)}+\xi_2^{(1)}$ is infinitesimal and 
$M_{\epsilon}$ is some bilinear operation on $\Lambda_1$ and $\Phi_1$, are the symmetries of 
equations \rf{meu2} iff 
\begin{eqnarray}
K_{\epsilon}(D\Lambda_1,\Phi_1)+ K_{\epsilon}(\Phi_1, D\Lambda_1)=DM_{\epsilon}(\Lambda_1,\Phi_1). 
\end{eqnarray}
One can easily show that $K_{\epsilon}(D\Lambda_1,\Phi_1)$ is always closed with respect to $D$; 
we need now to find such 
$M_{\epsilon}$ that $K_{\epsilon}(D\Lambda_1,\Phi_1)$ is exact. To show that such $M_{\epsilon}$ exists we 
need the following Proposition. \\
{\bf Proposition 2.1.} {\it The expressions 
\begin{eqnarray}
\label{hol}f_1(V,W)(z)=\frac{1}{2\pi i}\int_{C_{\epsilon,z}}dwV(w)W(z)+\frac{1}{2\pi i}
\int_{C_{\epsilon,z}}dwW(w)V(z),\\
\label{ahol}f_2(V,W)(z)=\frac{1}{2\pi i}\int_{C_{\epsilon,z}}d\b wV(w)W(z)+\frac{1}{2\pi i}
\int_{C_{\epsilon,z}}d\b wW(w)V(z)
\end{eqnarray} 
can be represented in the following form:
\begin{eqnarray}
f_i(V,W)(z)=\p \b g_i(V,W)(z)+\bp g_i(V,W)(z)
\end{eqnarray} 
for some operators $g_i$ and $\b g_i$, built from $(V,W)_k^{(r,s)}$ and their derivatives 
(here $\p\equiv\frac{\p}{\p z},\bp\equiv\frac{\p}{\p \b z}$).}\\
\hspace*{5mm}
The proof can be easily obtained using the Assumption 1 and comparing the coefficients 
$(V,W)_k^{(r+1,r)}$ and $(W,V)_k^{(r+1,r)}$ for \rf{hol}, and the coefficients 
$(V,W)_k^{(r,r+1)}$ and $(W,V)_k^{(r,r+1)}$ for \rf{ahol}.\\ 
{\bf Proposition 2.2.}
 {\it The expression 
\begin{eqnarray}
\frac{1}{2\pi i}\int_{C_{\epsilon,z}}\lambda^{(1)}(w)
d\rho^{(1)}(z)-\frac{1}{2\pi i}\int_{C_{\epsilon,z}}\rho^{(1)}(w)
d\lambda^{(1)}(z)
\end{eqnarray}
is always exact with respect to the de Rham differential for any bosonic operator valued 1-forms $\lambda^{(1)}$ and $\rho^{(1)}$.}\\
{\bf Proof.} Let's denote $\lambda^{(1)}\equiv  \lambda(z)dz-\b \lambda(z)d\b z$ and 
$\rho^{(1)}\equiv  \rho(z)dz-\b \rho(z)d\b z$. Then, showing that 
\begin{eqnarray}\label{rl}
\frac{1}{2\pi i}\int_{C_{\epsilon,z}}\lambda^{(1)}(w)(\p\b \rho(z)+\bp\rho(z))-
\frac{1}{2\pi i}\int_{C_{\epsilon,z}}\rho^{(1)}(w)(\p\b \lambda(z)+\bp \lambda(z))
\end{eqnarray}
reduces to sum $\p \b \alpha+\bp \alpha$ for some operators $\b \alpha$ and $\alpha$, 
we prove the Proposition 2.2.. 
Let's consider the first line in \rf{rl}. 
Recalling that the action of $\p \cdot$ and $\bp \cdot$ is equivalent to the action of Virasoro generators 
$[L_{-1},\cdot]$ and $[\b L_{-1},\cdot]$ correspondingly, the first term of \rf{rl} 
can be rewritten as follows:
\begin{eqnarray}
&&[L_{-1}, \frac{1}{2\pi i}\int_{C_{\epsilon,z}}\lambda^{(1)}(w)\b \rho(z)]+
[\b L_{-1}, \frac{1}{2\pi i}\int_{C_{\epsilon,z}}\lambda^{(1)}(w)\rho(z)]-\nonumber\\
&& \frac{1}{2\pi i}\int_{C_{\epsilon,z}}([L_{-1},\lambda(w)]dw-[L_{-1},\b \lambda(w)]d\b w)
\b \rho(z)-\nonumber\\
&&\frac{1}{2\pi i}\int_{C_{\epsilon,z}}([\b L_{-1},\lambda(w)]dw-[\b L_{-1},\b \lambda(w)]d\b w)
\rho(z).
\end{eqnarray}
We can see that the first line in the formula above 
is represented in the needed form, while the second and the third lines can be reexpressed:
\begin{eqnarray}\label{lr1}
&&\frac{1}{2\pi i}\int_{C_{\epsilon,z}}d\b w([\b L_{-1},\lambda(w)]+[L_{-1},\b \lambda(w)])
\b \rho(z)-\nonumber\\
&&\frac{1}{2\pi i}\int_{C_{\epsilon,z}}dw([\b L_{-1},\lambda(w)]+[ L_{-1},\b \lambda(w)]) \rho(z),
\end{eqnarray} 
using the fact that the integral of the total derivative vanishes. 
Let's compare this with the second line in \rf{rl}:
\begin{eqnarray}\label{lr2}
\frac{1}{2\pi i}\int_{C_{\epsilon,z}}\b \rho(w)d\b w([L_{-1},\b \lambda](z)+[\b L_{-1}, \lambda](z))-\nonumber\\
\frac{1}{2\pi i}\int_{C_{\epsilon,z}}\rho(w)d w([L_{-1},\b \lambda](z)+[\b L_{-1}, \lambda](z)).
\end{eqnarray} 
In order to see that the sum of \rf{lr1} and \rf{lr2} is equal to the sum $\p\b \beta +\bp \beta$ for some 
$\beta$, one needs to use Proposition 2.1. $\blacksquare$ \\
Let's consider the following type of transformation of $\phi_2^{(2)}$:
\begin{eqnarray}\label{symm}
\delta \phi_2^{(2)}(z)=-d(\xi_2^{(1)}(z)+\chi^{(1)}(\xi_1^{(1)},\phi_1^{(1)}) ) +
\frac{1}{2\pi i}\int_{C_{\epsilon, z}}\xi_1^{(1)}(w)\phi_1^{(2)}(z)
\end{eqnarray}
and $\psi_2^{(1)}$:
\begin{eqnarray}
&&\delta\psi_2^{(1)}(z)=d(\alpha^{(0)}_2(z)+\nu^{(0)}(\xi_1^{(1)},\psi_1^{(2)}))
-[Q,\xi_2^{(1)}+\chi^{(1)}(\xi_1^{(1)},\phi_1^{(2)})](z)+\nonumber\\
&&\eta_{\epsilon}^{(1)}(\xi_1^{(1)},\psi_1^{(1)})(z),
\end{eqnarray}
where  $\eta_{\epsilon}^{(1)}$ can be obtained from the following relation: 
\begin{eqnarray}\label{eta}
\frac{1}{2\pi i}\int_{C_{\epsilon,z}}\xi_1^{(1)}(w)
d\psi_1^{(1)}(z)-\frac{1}{2\pi i}\int_{C_{\epsilon,z}}\psi_1^{(1)}(w)
d\xi_1^{(1)}(z)=d\eta_{\epsilon}^{(1)}(\xi_1^{(1)},\psi_1^{(1)})(z),
\end{eqnarray}
$\chi^{(1)}$ and $\nu^{(0)}$ are some bilinear operations on operator-valued 1-forms and 2-forms. 
It is easy to see that equation \rf{me2} is invariant under such type of a transformation and thus the 
expression for $M_{\epsilon}$ is:
\begin{eqnarray}
&&M_{\epsilon}(\Lambda_1,\Phi_1)(z,\theta)=-D(\chi^{(1)}(\xi_1^{(1)},\phi_1^{(1)})+
\theta\nu^{(0)}(\xi_1^{(1)},\psi_1^{(2)}))+\nonumber\\
&&\frac{1}{2\pi i}\int_{C_{\epsilon, z}}\xi_1^{(1)}(w)\phi_1^{(2)}(z)+\theta 
\eta_{\epsilon}^{(1)}(\xi^{(1)},\psi^{(1)})(z).
\end{eqnarray}
However, the concrete choice of the bilinear operations $\chi^{(1)}, \nu^{(0)}$, and $\eta_{\epsilon}^{(1)}$ 
(it is defined up to some closed 1-form) is unclear 
in the general situation. So, here we have only the class of transformations under which 
\rf{meu1} and \rf{meu2} 
are invariant; in order to pick out a symmetry, which should be a part of a symmetry of more general 
equation \rf{genmc}, 
one should find the expressions for higher operations. 
\section{Deformation of BRST operator}  
\subsection{ OPE and Operator Equations}
\noindent Let's consider the following expressions for the components of the perturbation 2-forms from equations \rf{me1} and \rf{me2}: 
\begin{eqnarray}
\phi_1^{(2)}=V_1(z)dz\wedge d\b z, \quad \phi_2^{(2)}=(V_2(z)+\epsilon-terms)dz\wedge d\b z.
\end{eqnarray}
By $\epsilon-terms$ we mean the possible $\epsilon$-dependent terms from $\phi_2^{(2)}$. We say that 
the pair $V_1$ and $V_2$ is {\it of type (1,1)} if the following relations take place:
\begin{eqnarray}
&&1).\quad(L_{1+k}V_i)(z)=(\b {L}_{1+k}V_i)(z)=0, \quad k>0 \quad (i=1,2)\label{opetv}\\
&&(L_0V_i)(z)=(\b L_0V_i)(z);\nonumber\\
&&\label{opevv} 2). \quad V_1(z)V_1(z')=
\sum_{s=-\infty}^{2}\sum_{r=-\infty}^{2}\sum^{\infty}_{i=0}(V_1,V_1)^{(r,s)}_i(z')(z-z')^{-r}(\b z-\b z')^{-s}
\nonumber\\
&&(\alpha'\ln|(z-z')/\mu|)^i,\\
&& (L_1V_1)(z)V_1(z')=
\sum_{s=-\infty}^{2}\sum_{r=-\infty}^{1}\sum^{\infty}_{i=0}(L_1V_1,V_1)^{(r,s)}_i(z')(z-z')^{-r}(\b z-\b z')^{-s}
\nonumber\\
&&(\alpha'\ln|(z-z')/\mu|)^i,\\
&& (\b L_1V_1)(z)V_1(z')=
\sum_{s=-\infty}^{1}\sum_{r=-\infty}^{2}\sum^{\infty}_{i=0}(\b L_1V_1,V_1)^{(r,s)}_i(z')(z-z')^{-r}(\b z-\b z')^{-s}
\nonumber\\
&&(\alpha'\ln|(z-z')/\mu|)^i,
\end{eqnarray}
where $L_n$ are Virasoro operators.\\ 
{\bf Proposition 3.1.} {\it The coefficients in the OPE (\ref{opevv}) are not independent and the following relations 
take place:
\begin{eqnarray}
\label{vrel}
&&(V_1,V_1)_i^{(1,2)}=\half\partial (V_1,V_1)_i^{(2,2)},\quad
(V,V)_i^{(2,1)}=\half \bar{\partial} (V,V)_i^{(2,2)},\\
&&(V,V)_i^{(1,0)}=\half (\bar{\partial}(V,V)_i^{(1,1)}-
\half \partial\bar{\partial}^2 (V,V)_i^{(2,2)}+\partial (V,V)_i^{(2,0)}),\nonumber\\
&&(V,V)_i^{(0,1)}=\half (\partial (V,V)_i^{(1,1)}-
\half \partial^2\bar{\partial}(V,V)_i^{(2,2)}+\bar{\partial} (V,V)_i^{(0,2)}).\nonumber
\end{eqnarray}}\\
To prove this one should expand the OPE $V_1(z_1)V_1(z_2)$ around $z_1$ and $z_2$ and compare the coefficients.
\\
\hspace*{5mm} 
Now let's recall the definition of the BRST operator in CFT \cite{brst}:
\begin{eqnarray}
\label{BRST}
&&Q_B=\frac{1}{2\pi i}\oint\mathcal{J_B},\quad \mathcal{J_B}=j_Bdz-\tilde{j_B}d\bar{z},\\
&&j_B=cT+:bc\partial c:+\frac{3}{2}\partial^2 c, \quad \tilde{j}_B=\tilde{c}\tilde{T}
+:\tilde{b}\tilde{c}
\bar{\partial} \tilde{c}:+\frac{3}{2}\bar{\partial}^2 \tilde{c},\nonumber
\end{eqnarray}
where $b$ and $c$ $(\t b$ and $\t c)$ are the primary operators with conformal weights (2,0) and (-1,0) 
((0,2) and (0,-1)) correspondingly with the operator 
product $c(z)b(w)\sim \frac{1}{z-w}$ ($\t c(z)\t b(w)\sim \frac{1}{\b z-\b w}$).\\
 The BRST current obeys the following relation:
\begin{eqnarray}
(b_0\mathcal{J_B})(z)-(\t b_0\mathcal{J_B})=j_{g}(z)dz+\t j_{g}(z)d\b z,
\end{eqnarray}
where $j_g=:bc:$ and $\t j_g=:\t b\t c:$.\\ 
This motivates us to formulate the constraints we need to put on a deformation of BRST current.\\  
\hspace*{5mm}We say that the operator-valued 1-form $\psi^{(1)}$ is a deformation of the 
BRST operator if:\\ 
1). $\psi^{(1)}$ is of ghost number 1 and of the first order in $c$ and $\t c$ and their derivatives. \\
2). $\psi^{(1)}$ obeys the projection relation:
\begin{eqnarray}
(b_0\psi^{(1)})(z)-(\t b_0 \psi^{(1)})(z)=0.
\end{eqnarray}
These constraints allow to write a {\it (1,1) ansatz} for $\psi_1^{(2)}$ and $\psi_2^{(2)}$ corresponding 
to the (1,1) type of the perturbation:  
\begin{eqnarray}
&&\psi_{1}^{(1)}=(cY_1-\t c\b Z_1 -(\p c+\bp \tc)\b W_1 -1/2\bp^2 \tc \b U_1)d\b z
+\nonumber\\
&&(cZ_1-\tc \b Y_1+(\p c+\bp \tc)W_1+1/2\p^2 c U_1)dz
\end{eqnarray}
and 
\begin{eqnarray}
&&\psi_{2}^{(1)}=(cY_2-\t c\b Z_2 -(\p c+\bp \tc)\b W_2 -1/2\bp^2 \tc \b U_2)d\b z
+\nonumber\\
&&(cZ_2-\tc \b Y_2+(\p c+\bp \tc)W_2+1/2\p^2 c U_2)dz+\epsilon-terms,
\end{eqnarray}
such that the OPEs between ``dilatonic'' terms $U_1$, $\b U_1$, and $V_1$ have the following form:
\begin{eqnarray} 
&& U_1(z)V_1(z')=
\sum_{s=-\infty}^{1}\sum_{r=-\infty}^{1}\sum^{\infty}_{i=0}(U_1,V_1)^{(r,s)}_i(z')(z-z')^{-r}
(\b z-\b z')^{-s}
\nonumber\\
&&(\alpha'\ln|(z-z')/\mu|)^i,\\
&& \b U_1(z)V_1(z')=
\sum_{s=-\infty}^{1}\sum_{r=-\infty}^{1}\sum^{\infty}_{i=0}(\b U_1,V_1)^{(r,s)}_i(z')(z-z')^{-r}(\b z-\b z')^{-s}
\nonumber\\
&&(\alpha'\ln|(z-z')/\mu|)^i.
\end{eqnarray}
The absence of higher derivatives of $c$ and $\tc$ in the ansatz above is motivated by the fact that the 
BRST operator acting on $V_i$ $(i=1,2)$, and the operation $K_{\epsilon}$ will not contain monomials 
with a number of derivatives of $c$ and $\t c$ higher than two, 
so if we include such type of terms in $\psi^{(1)}_i$, they will interact among themselves and will not put any 
constraints on perturbation. The OPEs between $U$- and $V$-operators are chosen because of dimensional reasons. 
\\
\hspace*{5mm} 
After these preparations, we are ready to formulate the following proposition.\\ 
{\bf Proposition 3.2.} {\it For the pair $\phi_1^{(2)}$ and $\phi_2^{(2)}$ of type (1,1)  
and (1,1) ansatz for $\psi_1^{(2)}$ and $\psi_2^{(2)}$ we have:\\  
1. Equation \rf{me1} puts the following constraint on $V_1, U_1$ and $\b U_1$:
\begin{eqnarray}\label{m1}
&&(L_0V_1)(z)-V_1(z)-1/2L_{-1}L_1V_1-1/2\b L_{-1}\b L_1V_1-\nonumber\\
&&1/2L_{-1}\b L_{-1} (U_1+\b U_1)=0.
\end{eqnarray}
2. Putting $Z_1, \b Z_1=0$ and using the relations from point 1, one obtains the following constraint on 
 $V_i, U_i$ and $\b U_i$ (i=1,2) from \rf{me2}:
\begin{eqnarray}\label{m2}
&&(L_0V_2)(z)-V_2(z)-1/2(V_1,V_1)_0^{(1,1)}(z)-1/2
( L_1V_1+\b L_{-1} U_1,V_1)_0^{(0,1)}(z)-\nonumber\\ 
&& 1/2(\b L_1V_1+L_{-1} \b U_1,V_1)_0^{(1,0)}(z)+L_{-1}\b W_2(z)+ \b L_{-1} W_2(z)=0,
\end{eqnarray}
where 
\begin{eqnarray}
&&\b W_2(z)=-1/2((L_1 V_2)(z)-(L_1V_1+\b L_{-1} U_1,V_1)_0^{(1,1)}(z)-\nonumber\\
&&(\b L_1V_1+L_{-1} \b U_1,V_1)_0^{(2,0)}(z)+(U_1,V_1)_0^{(1,0)}(z)+\b L_{-1}U_2(z))\nonumber\\
&&W_2(z)=-1/2((\b L_1 V_2)(z)-(\b L_1V_1+L_{-1} \b U_1,V_1)_0^{(1,1)}(z)-\nonumber\\
&&(L_1V_1+\b L_{-1} U_1,V_1)_0^{(0,2)}(z)+(\b U_1,V_1)_0^{(0,1)}(z)+ L_{-1}\b U_2(z)).\nonumber
\end{eqnarray}} 
The Proof is given in Appendix A. 
\\
\hspace*{5mm} 
We draw an attention that there is an ambiguity related to unconstrained $Z^h_1$ and $Z^a_1$ terms. 
We will describe the way how one can get rid of them in the next subsection, substantiating the 
$Z_1, \b Z_1=0$ condition in point 2 of Proposition 3.2.
\\
\hspace*{5mm} 
We note, that there is possibly an infinite set of equations for renormalization $\epsilon$-dependent terms. 
In this article, we don't consider them, because they do not put any constraints on the basic part of the 
perturbation, i.e. $V_1$ and $V_2$.
\subsection{ Master Equation}
\noindent For any operator $\phi^{(0)}$ of ghost number 2, we define 1-form and 2-form by means of the b-ghost:
\begin{eqnarray}
&&\label{d1}\phi_1^{(1)}\equiv dz[b_{-1},\phi_1^{(0)}]+d\b z[\t b_{-1},\phi_1^{(0)}],\\
&&\label{d2}\phi_1^{(2)}\equiv dz\wedge d\b z[b_{-1}, [\t b_{-1},\phi_1^{(0)}]]. 
\end{eqnarray}
In the case when $\psi_1^{(1)}$ is equal to $\phi_1^{(1)}$ for some $\phi_1^{(0)}$, the equation \rf{me1}
can be reexpressed as follows:
\begin{eqnarray}\label{phi10}
[b_{-1},[\t b_{-1},[Q_B,\phi_1^{(0)}]]]=0.
\end{eqnarray}
If we suppose that $\psi_1^{(1)}=\phi_1^{(1)}$ for some $\phi_1^{(0)}$, then we obtain the 
condition $Z_1, \b Z_1=0$, since projection \rf{d2} can't afford such type of terms in $\phi_1^{(1)}$ while 
other terms, considered in (1,1) ansatz, survive. This allows thinking about the following: 
whether, in the case $Q=Q_B$, equation \rf{me2} can be reformulated also 
in terms of modes of $b, \t b$ -ghosts and $\phi^{(0)}$ operator only. It appears that the answer is positive, 
that is, the following proposition holds:\\
{\bf Proposition 3.3.}
{\it The equation 
\begin{eqnarray}\label{phi20}
[b_{-1},[\t b_{-1},[Q_B,\phi_2^{(0)}]+\frac{1}{4\pi i}\int_{C_{\epsilon,z}} \phi_1^{(1)}(w)\phi_1^{(0)}(z)]]=0
\end{eqnarray}
leads to the exactness of the following 2-form  
\begin{eqnarray}
[Q_B,\phi_2^{(2)}](z)+\frac{1}{2\pi i}\int_{C_{\epsilon,z}} \phi_1^{(1)}(w)\phi_1^{(2)}(z).
\end{eqnarray}}
{\bf Proof}. 
We use the approach, similar to the one used in the proof of Proposition 2.2. 
Really, if we show that the operation 
\begin{eqnarray}\label{bfi0}
[b_{-1},[\t b_{-1},\frac{1}{4\pi i}\int_{C_{\epsilon,z}} (dw[b_{-1},\phi_1^{(0)}](w)+
d\b w[\t b_{-1},\phi_1^{(0)}](w))\phi_1^{(0)}(z)]]
\end{eqnarray}
reduces to the sum $\p\b \alpha +\bp \alpha$ for some $\alpha$ and $\b \alpha$, we will get the desired result. 
Pushing $b_{-1}$, $\t b_{-1}$ through the integration in \rf{bfi0}, we can 
rewrite the expression above as:
\begin{eqnarray}\label{pbp}
&&\frac{1}{4\pi i}\int_{C_{\epsilon,z}} (dw[b_{-1},\phi_1^{(0)}](w)+d\b w[\t b_{-1},\phi_1^{(0)}](w))
[b_{-1},[\t b_{-1}, \phi_1^{(0)}(z)]]+\nonumber\\
&&\frac{1}{4\pi i}\int_{C_{\epsilon,z}} dw[b_{-1},[\t b_{-1}, \phi_1^{(0)}(w)]][b_{-1},\phi_1^{(0)}](z)+\nonumber\\
&&\frac{1}{4\pi i}\int_{C_{\epsilon,z}}d\b w[b_{-1},[\t b_{-1}, \phi_1^{(0)}(w)]][\t b_{-1},\phi_1^{(0)}](z)
\end{eqnarray}
Using the Proposition 2.1., we 
prove the Proposition 3.3.
$\blacksquare$\\ 
\hspace*{5mm}So, we got the system of two equations 
\begin{eqnarray}
&&\label{ME1}[b_{-1},[\t b_{-1},[Q_B,\phi_1^{(0)}]=0,\\
&&\label{ME2}[b_{-1},[\t b_{-1},[Q_B,\phi_2^{(0)}]+\frac{1}{4\pi i}\int_{C_{\epsilon,z}} \phi_1^{(1)}(w)\phi_1^{(0)}(z)]]=0,
\end{eqnarray}
such that  \rf{me1} and \rf{me2} 
with needed constraint $Z_1, \b Z_1=0$ are its consequences. In the following we will call the system of equations \rf{ME1} and 
\rf{ME2} {\it the Master Equation}, since we hope they could be unified in one equation 
\begin{eqnarray}\label{ME}
[b_{-1},[\t b_{-1},[Q_B,\phi^{(0)}]+\frac{1}{4\pi i}\int_{C_{\epsilon,z}} \phi^{(1)}(w)\phi^{(0)}(z)+...]]=0,
\end{eqnarray}  
where dots mean higher operations in $\phi^{(0)}$ such that \rf{ME1} and 
\rf{ME2} are the first two nontrivial equations with respect to the expansion $\phi^{(0)}=t\phi_1^{(0)}+t^2\phi_2^{(0)}+
O(t^3)$. Therefore, this equation can be treated as the generalized Maurer-Cartan equation with a projection.  
One can suspect that the symmetries of equation \rf{ME} can be 
expressed as follows:
\begin{eqnarray}
\delta \phi^{(0)}=[Q_B,\xi^{(0)}]+N_{\epsilon}(\xi^{(0)},\phi^{(0)})+... 
\end{eqnarray} 
for some operator $\xi^{(0)}$ of ghost number 1, where dots again stands for higher operations.
\\
\hspace*{5mm} 
Equations \rf{ME1} and \rf{ME2} possess a family of transformations, which leaves them invariant. 
Really, we consider $\xi_i^{(0)}$ $(i=1,2)$ such that $[b_{-1},[\t b_{-1}, \xi_i^{(0)}]]=0$. Then the transformations 
\begin{eqnarray}\label{masym}
&&\delta \phi_1^{(0)}=[Q_B,\xi_1^{(0)}],\nonumber\\
&&\delta \phi_2^{(0)}=[Q_B,\xi_2^{(0)}+\rho^{(0)}(\xi_1^{(0)},\phi_1^{(0)})]+
\frac{1}{4\pi i}\int_{C_{\epsilon,z}} \xi_1^{(1)}(w)\phi_1^{(0)}(z)+\nonumber\\
&&\frac{1}{4\pi i}\int_{C_{\epsilon,z}} \phi_1^{(1)}(w)\xi_1^{(0)}(z),
\end{eqnarray} 
where $\xi_1^{(1)}=dz[b_{-1},\xi_1^{(0)}]+d\b z[\t b_{-1},\xi_1^{(0)}]$ and $\rho^{(0)}$ is some bilinear 
operation, leave 
\rf{ME1} and \rf{ME2} unchanged. Therefore, the symmetries of global equation \rf{ME} belong to this family at the second order 
of $t$-expansion with $N_{\epsilon}$ given by 
\begin{eqnarray}
&&N_{\epsilon}(\xi^{(0)},\phi^{(0)})=
[Q_B,\rho^{(0)}(\xi^{(0)},\phi^{(0)})]+\nonumber\\
&&\frac{1}{4\pi i}\int_{C_{\epsilon,z}} \xi^{(1)}(w)\phi^{(0)}(z)+\frac{1}{4\pi i}\int_{C_{\epsilon,z}} \phi^{(1)}(w)\xi^{(0)}(z).
\end{eqnarray}
Applying $b_{-1}\t b_{-1}$ to both sides of expression \rf{masym} and using 
Proposition 2.1, we get the formulae for the transformation of $\phi_i^{(2)}$, 
which are equivalent to \rf{symm}, as one should expect.
\\
\hspace*{5mm} 
Now, let's find the relations between \rf{m1} and \rf{m2} and similar relations arising from 
Master Equation. The ansatz for $\phi_1^{(0)}$, which corresponds 
to (1,1) ansatz of $\psi_1^{(1)}=\phi_1^{(1)}$, is:
\begin{eqnarray}\label{aphi1}
&&\phi_1^{(0)}=\nonumber\\
&&\tc c V_1-\t c(\p c+ \bp \tc)\b W_1-1/2\t c\bp^2 \t c\b U_1+
c(\p c+ \bp \tc)W_1+1/2c\p^2 cU_1.
\end{eqnarray}
Then we take a similar ansatz for $\phi_2^{(0)}$:
\begin{eqnarray}\label{aphi2}
&&\phi_2^{(0)}=\tc c V_2-\t c(\p c+ \bp \tc)\b \mathcal{W}_2-1/2\t c\bp^2 \t c\b \mathcal{U}_2+
c(\p c+ \bp \tc)\mathcal{W}_2+1/2c\p^2 c\mathcal{U}_2+\nonumber\\
&&\epsilon-terms.
\end{eqnarray}
It is worth noting that the ghost structure of $\phi_i^{(0)}$ $(i=1,2)$ can be obtained 
from the two conditions:\\
1) $\phi_i^{(0)}$ are of the ghost number 2 and of the second order in $c$ and $\t c$ 
and their derivatives;\\  
2) $(b_0\phi_i^{(0)})-(\t b_0\phi_i^{(0)})$=0. \\
The absence of higher derivative terms of $c$ and $\t c$ can be substantiated as in the case of (1,1) ansatz for 
$\psi_i^{(1)}$ in the previous subsection, i.e. if we include higher derivative terms, then the  
Master equation will create only the relations in which they interact among themselves and do 
not put any constraints on $V_1$ and the dilatonic terms $U_1$ and $\b U_1$.\\ 
Using the proposed ansatz, we are ready to formulate the following proposition.\\
{\bf Proposition 3.4.} 
{\it The ansatz given in \rf{aphi1} together with \rf{ME1} gives the 
familiar relation \rf {me1}:
\begin{eqnarray}\label{m10}
&&(L_0V_1)(z)-V_1(z)-1/2L_{-1}L_1V_1-1/2\b L_{-1}\b L_1V_1-\nonumber\\
&&1/2L_{-1}\b L_{-1} (U_1+\b U_1)=0\nonumber
\end{eqnarray}
and ansatz from \rf{aphi2} and equation \rf{ME2} leads to the relation:
\begin{eqnarray}\label{m20}
&&(L_0V_2)(z)-V_2(z)-1/2(V_1,V_1)_0^{(1,1)}(z)-1/4
( L_1V_1+\b L_{-1} U_1,V_1)_0^{(0,1)}(z)+\nonumber\\
&&1/4
( V_1,L_1V_1+\b L_{-1} U_1)_0^{(0,1)}(z)
-1/4(\b L_1V_1+L_{-1} \b U_1,V_1)_0^{(1,0)}(z)+\nonumber\\
&&1/4(V_1,\b L_1V_1+L_{-1} \b U_1)_0^{(1,0)}(z)+L_{-1}\b \mathcal{W}_2(z)+ \b L_{-1} \mathcal{W}_2(z)=0,
\end{eqnarray}
where 
\begin{eqnarray}
&&\b \mathcal{W}_2(z)=-1/2((L_1 V_2)(z)-1/2(V_1,V_1)_0^{(2,1)}-1/2(L_1V_1+\b L_{-1} U_1,V_1)_0^{(1,1)}(z)-\nonumber\\
&&1/2(\b L_1V_1+L_{-1} \b U_1,V_1)_0^{(2,0)}(z)+1/2(U_1,V_1)_0^{(1,0)}(z)-
1/2(V_1,U_1)_0^{(1,0)}(z)+\nonumber\\
&&\b L_{-1}\mathcal{U}_2(z)),\nonumber\\
&&\mathcal{W}_2(z)=-1/2((\b L_1 V_2)(z)-1/2(V_1,V_1)_0^{(1,2)}-1/2(\b L_1V_1+L_{-1} \b U_1,V_1)_0^{(1,1)}(z)-\nonumber\\
&&1/2(L_1V_1+\b L_{-1} U_1,V_1)_0^{(0,2)}(z)+1/2(\b U_1,V_1)_0^{(0,1)}(z)-
1/2(V_1,\b U_1)_0^{(0,1)}(z)+\nonumber \\
&&L_{-1}\b \mathcal{U}_2(z)).\nonumber
\end{eqnarray}
For 
\begin{eqnarray}
\mathcal{U}_2=U_2+1/4(V_1,V_1)_0^{(2,2)}+1/2(U_1,V_1)_0^{(1,1)}-1/2(L_1V_1+\b L_{-1}U_1,V_1)_0^{(1,2)}\nonumber\\
\b \mathcal{U}_2=\b U_2+1/4(V_1,V_1)_0^{(2,2)}+1/2(\b U_1,V_1)_0^{(1,1)}-1/2(\b L_1V_1+L_{-1}\b U_1,V_1)_0^{(2,1)}
\nonumber
\end{eqnarray}
equation \rf{m20} is equivalent to \rf{m2}.}\\
The proof is similar to that in the Proposition 3.2 from the previous subsection. 
So, we find that the constraints on $V_i, U_i$ and $\b U_i$ arising both from \rf{me1} and \rf{me2} with the 
condition $Z_1, \b Z_1 =0$ are equivalent to the ones obtained from \rf{ME1} and \rf{ME2}. 
\section{Example: First Order Theory}
\subsection {Master Equation and the First Order Theory}
\noindent Let's consider the 2d CFT with the action \cite {kap}-\cite{nekrasov}:  
\begin{eqnarray}\label{free}
S_0=\frac{1}{2\pi\alpha'}\int d^2 z (p_i\bar{\partial} X^{i}+
p_{\bar{i}}{\partial} X^{\bar{i}}),
\end{eqnarray}
(known as (curved) beta-gamma system, 
by the simple analogy with superconformal ghosts \cite{pol}), where $i=1,... ,D/2$ ($D$-even) 
and the momentum $p,\bar{p}$-fields are the $(1,0)$- and $(0,1)$-forms correspondingly, 
while the co-ordinates $X$ and $\bar{X}$ are scalars with the weights $(0,0)$. 
The equations of motion, following from the Lagrangian \rf{free}, provide the 
nontrivial operator product expansions (OPE):
\begin{eqnarray}
&&X^{i}(z_1)p_{j}(z_2)\sim\frac{\alpha'\delta^{i}_{j}}{z_1-z_2}, \quad
X^{\bar{i}}(\bar{z}_1)p_{\bar{j}}(\bar{z}_2)
\sim\frac{\alpha'\delta^{\bar{i}}_{\bar{j}}}
{\bar{z}_1-\bar{z}_2}
\end{eqnarray}
(it is convenient to keep here explicit $\alpha'$-dependence) 
and there are no singular contractions between the $X$- and $p$-fields
themselves.
\\
\hspace*{5mm} 
The components of the energy-momentum tensor are:
\begin{eqnarray}
T=-(\alpha')^{-1}p_i\partial X^i, \quad \tilde{T}=-(\alpha')^{-1}p_{\bar{i}}\bar{\partial}X^{\bar{i}}.
\end{eqnarray}\\
The action $S_0$, perturbed by the following operator 
\begin{eqnarray}
&&\phi^{(2)}=dz\wedge d\b z\alpha'^{-1}(g^{i\bar{j}}(X,\b X)p_i p_{\bar{j}}+\bar{\mu}^{\bar{j}}_i(X,\b X)\partial X^i p_{\bar{j}}+
\mu_{\bar{i}}^j(X,\b X)\bar{\partial}X^{\bar{i}} p_j + \nonumber\\
&&b_{i\bar{j}}(X,\b X)\partial X^i
\bar{\partial} X^{\bar{j}}),
\end{eqnarray} 
is called the first order nonlinear sigma model \cite{lmz}, since after integration over $p_i$ and 
$p_{\bar{j}}$ variables, we arrive to the usual second order sigma model. 
In the following, for the simplicity of calculations, 
we will consider the hypersurface in the whole given space of perturbations parametrised by 
$g,\mu,\b \mu$, and $b$, restricting our perturbation to 
\begin{eqnarray}
\phi_g^{(2)}=dz\wedge d\b z\alpha'^{-1}g^{i\bar{j}}(X,\b X)p_i p_{\bar{j}}.
\end{eqnarray}
Expanding $g^{i\bar{j}}=tg_1^{i\bar{j}}+t^2g_2^{i\bar{j}}+O(t^3)$, adding 
possible $\epsilon$-dependent terms, 
and applying the Master Equation we have the following proposition:\\ 
{\bf Proposition 4.1.} {\it The constraints \rf{m1} and \rf{m2} applied to  
\begin{eqnarray}\label{fopert}
V_{1}=\alpha'^{-1}g_1^{i\bar{j}}(X,\bar{X})p_i p_{\bar{j}}, \quad 
V_{2}=\alpha'^{-1}g_2^{i\bar{j}}(X,\bar{X})p_i p_{\bar{j}},
\end{eqnarray}
where $U_i\equiv f_i(X,\b X)$ and $\b U_i\equiv \b f_i(X,\b X) $ (i=1,2) are just the functions of $X^i,X^{\b j}$, 
give the following equations at zero order 
in $\alpha'$:  
\begin{eqnarray}\label{foit}
&&\p_{\bar{p}}\p_{\bar{l}}g_1^{\bar{l}k}=0, \quad \p_{p}\p_{l}g_1^{\bar{k}l}=0,\nonumber\\
&&\p_i\p_{\bar{k}}\Phi_{0, s}=0, \quad (s=1,2),\nonumber\\
&&2g_1^{r\bar{l}}\p_r\p_{\bar{l}}g_1^{i\bar{k}}-2\p_r g_1^{i\bar{p}}\p_{\bar{p}}g_1^{r\bar{k}}-
g_1^{i\bar{l}}\p_{\bar{l}}\p_sg_1^{s\bar{k}}-g_1^{r\bar{k}}\p_r \p_{\bar{j}}g_1^{\bar{j}i}+\nonumber\\
&&\p_rg_1^{i\bar{k}}\p_{\bar{j}}g_1^{\bar{j}r}+\p_{\bar{p}}g_1^{\bar{k}i}\p_n g_1^{n\bar{p}}=0,\nonumber\\
&&\p_{\bar{p}}(\p_{\bar{l}}g_2^{\bar{l}k}-\p_{\bar{l}}(\Phi_{0, 1})g_1^{\bar{l}k})=0, 
\quad \p_{p}(\p_{l}g_2^{\bar{k}l}-2\p_{l}(\Phi_{0,1})g_1^{\bar{k}l})=0,
\end{eqnarray}
where $\Phi_{0,s}=1/2(f_s(X,\b X)+\b f_s(X,\b X))$.}\\
The proof can be easily obtained by simple substitution \rf{fopert} in \rf{m1} and \rf{m2}.\\
Thus we have proceeded from the operator equations \rf{m1} and \rf{m2} to the 
field equations expanded with respect to the formal parameter, i.e. \rf{foit}. 
As one might suspect, these field equations coincide with Einstein equations for an appropriate choice 
of metric, Kalb-Ramond field, and a dilaton.\\ 
{\bf Proposition 4.2.} 
{\it The Einstein equations 
\begin{eqnarray}
&&R^{\mu\nu}={1\over 4} H^{\mu\lambda\rho}H^{\nu}_{\lambda\rho}-2\nabla^{\mu}
\nabla^{\nu}\Phi,\\
&&\nabla_{\mu}H^{\mu\nu\rho}-2(\nabla_{\lambda}\Phi)H^{\lambda\nu\rho}=0,
\end{eqnarray}
where metric, B-field, and a dilaton are expressed as follows: 
\begin{eqnarray}
G_{i\bar{k}}=g_{i\bar{k}}, \quad B_{i\bar{k}}=-g_{i\bar{k}}, \quad \Phi=\log\sqrt{g}+\Phi_0,
\end{eqnarray}
are equivalent to the following system:
\begin{eqnarray}\label{feinst}
&&\label{harm}\p_i\p_{\bar{k}}\Phi_0=0,\label{vol}\\
&&\p_{\bar{p}}d^{\Phi_0}_{\bar{l}}g^{\bar{l}k}=0, \quad \p_{p}d^{\Phi_0}_{l}g^{\bar{k}l}=0,\label{hom}\\
&&2g^{r\bar{l}}\p_r\p_{\bar{l}}g^{i\bar{k}}-2\p_r g^{i\bar{p}}\p_{\bar{p}}g^{r\bar{k}}-
g^{i\bar{l}}\p_{\bar{l}}d^U_sg^{s\bar{k}}-g^{r\bar{k}}\p_r d^U_{\bar{j}}g^{\bar{j}i}+\nonumber\\
&&\p_rg^{i\bar{k}}d^U_{\bar{j}}g^{\bar{j}r}+\p_{\bar{p}}g^{\bar{k}i}d^U_n g^{n\bar{p}}=0,
\label{inhom}
\end{eqnarray}
where 
\begin{eqnarray}
&&d^{\Phi_0}_ig^{i\bar{j}}\equiv \p_i g^{i\bar{j}}-2\p_i\Phi_0 g^{i\bar{j}},\nonumber\\
&&d^{\Phi_0}_{\b i}g^{\b i j}\equiv \p_{\b i} g^{j\bar{i}}-2\p_{\b i} \Phi_0 g^{j\bar{i}}.
\end{eqnarray}
}\\
The proof is given in Appendix B.\\
\hspace*{5mm}The equations \rf{foit} can be obtained from \rf{vol}-\rf{inhom} by means of substitution:
$g^{i\b j}=tg_1^{i\b j}+t^2g_2^{i\b j}+O(t^3)$ and $\Phi_0=t\Phi_{0,1}+t^2\Phi_{0,2}+O(t^3)$.\\ 
\hspace*{5mm}From the equation \rf{harm} one can see, that $\Phi_0$ is the sum of holomorphic and 
antiholomorphic functions and gives contribution to the dilaton. Let's look on the consequences.
One can perturb the BRST operator with the dilatonic term:
\begin{eqnarray}
&&Q_B^{\Phi_0}=Q_B+\oint dzc(\partial X^i \partial X^j\partial_i\partial_j\Phi_0+
\partial^2 X^i\partial_i\Phi_0)-\nonumber\\
&&\oint d\bar{z}\tilde{c}(\bar{\partial}X^{\bar{i}}\bar{\partial}X^{\bar{j}}
\partial_{\bar{i}}\partial_{\bar{j}}\Phi_0+
\bar{\partial}^2X^{\bar{i}}\partial_{\bar{i}}\Phi_0).
\end{eqnarray}
If we put the condition that $Q^{\Phi_0}$ is conserved in the unperturbed theory, 
this immediately leads to $\p_i\p_{\b k}\Phi_0=0$. 
One can easily verify that the Master equation with $Q=Q_B^{\Phi_0}$ leads to the the Einstein equations above 
with the dilaton
\begin{eqnarray}
\Phi=\log\sqrt{g}+\Phi_0
\end{eqnarray}
up to the second order in $t$-expansion: $g^{i\b j}=tg_1^{i\b j}+t^2g_2^{i\b j}+O(t^3)$.
\subsection{Symmetries in the First Order Theory}
\noindent Let's consider 1-forms 
\begin{eqnarray}
\xi_l^{(1)}=\alpha'^{-1}(v_l^i(X,\b X)p_i dz-\b v_l(X,\b X)^{\b i}p_{\b i} d{\b z}) \quad (l=1,2).
\end{eqnarray}
We consider the case when $v_l^i$ are holomorphic functions on the target space and 
$v_l^{\b i}$ are antiholomorphic ones. Let's recall that according to \rf{symm}, the symmetry transformations are 
given by the following expressions:
\begin{eqnarray}
&&\delta \phi_1^{(2)}(z)=-d\xi_1^{(1)}(z),\\
&&\delta \phi_2^{(2)}(z)=-d(\xi_2^{(1)}(z)+\chi^{(1)}(\xi_1^{(1)},\phi_1^{(1)}) ) +
\frac{1}{2\pi i}\int_{C_{\epsilon, z}}\xi_1^{(1)}(w)\phi_1^{(2)}(z).
\end{eqnarray}
Let's put $\chi^{(1)}$ equal to zero, then $\delta g_l^{i\b k}$ ($l=1,2$) in the limit $\alpha'\to 0$
are:
\begin{eqnarray}
&&\delta_v g_1^{\bar{i}j}=0,\nonumber\\
&&\delta_v g_2^{\bar{i}j}= v_1^k\partial_k g_1^{\bar{i}j}+
\bar{v}_1^{\bar{k}}\partial_{\bar{k}}g_1^{\bar{i}j}-
\partial_{\bar{k}}\bar{v}_1^{\bar{i}}g_1^{\bar{k}j}-
\partial_{k}v_1^j g_1^{\bar{i}k}.
\end{eqnarray}
Correspondingly, remembering that the transformations 
of $\psi_i^{(2)}(z)$ are given by the formulas
\begin{eqnarray}
&&\delta\psi_1^{(1)}(z)=d\alpha^{(0)}_1(z)-[Q,\xi_1^{(1)}](z),\nonumber\\
&&\delta\psi_2^{(1)}(z)=d(\alpha^{(0)}_2(z)+\nu^{(0)}(\xi_1^{(1)},\psi_1^{(2)}))
-[Q,\xi_2^{(1)}](z)+\nonumber\\
&&\eta_{\epsilon}^{(1)}(\xi_1^{(1)},\psi_1^{(1)})(z),
\end{eqnarray}
and putting  $\eta_{\epsilon}^{(1)}(\xi_1^{(1)},\psi_1^{(1)})(z)\equiv \int_{C_{\epsilon,z}}\xi_1^{(1)}(w)
\psi_1^{(1)}(z)$, we find the following transformations for the dilaton: 
\begin{eqnarray}
&&\Phi_{0,1} \to \Phi_{0,1} +\p_iv_1^i+\p_{\b i}v_1^{\b i},\nonumber\\
&&\Phi_{0,2} \to \Phi_{0,2} +\p_iv_2^i+\p_{\b i}v_2^{\b i}+v_1^k\p_k\Phi_{0,1}+
v_1^{\b k}\p_{\b k}\Phi_{0,1}.\nonumber
\end{eqnarray}
If $v_l^i$ are not holomorphic, the symmetry leads to the change of the structure of the 
perturbing operator, that's why we don't consider such transformations. 
However, adding all possible perturbing terms
\begin{eqnarray}
V_{gen}=\alpha'^{-1}(g^{i\bar{j}}p_i p_{\bar{j}}+\bar{\mu}^{\bar{j}}_i\partial X^i p_{\bar{j}}+
\mu_{\bar{i}}^j\bar{\partial} X^{\bar{i}} p_j + b_{i\bar{j}}\partial X^i
\bar{\partial} X^{\bar{j}}),
\end{eqnarray}
one can consider the possibility of non(anti)holomorphic operators as the generators of symmetries.  
Also, in this case the additional type of symmetry, generated by 1-forms 
\begin{eqnarray}
\alpha'^{-1}(\omega_i\p X^i dz-\b \omega_{\b i}\bp X^{\b i} d\b z),
\end{eqnarray}
can appear. 
In the case of holomorphic $\omega_i$ and antiholomorphic 
$\omega_{\b i}$, it was already outlined in \cite{lmz}. \\
\subsection{Algebraic Structure of Einstein Equations for First Order Theory}
\noindent 
Equations \rf{feinst}-\rf{inhom} have a very interesting algebraic structure in the target space. 
One can rewrite the combination $g^{i\bar{j}}\partial_{i}\partial_{\bar{j}}g^{k\bar{l}}-
\partial_{i}g^{k\bar{j}}\partial_{\bar j}g^{i\bar{l}}$ which is present in \rf{feinst}
as follows:
\begin{eqnarray}\label{cond1alg}
&&g^{i\bar{j}}\partial_{i}\partial_{\bar{j}}g^{k\bar{l}}-
\partial_{i}g^{k\bar{j}}\partial_{\bar j}g^{i\bar{l}}=
\nonumber
\\
&&=\sum_{I,I'}\left((\mathcal{U}_{I}^{i}\partial_{i}\mathcal{U}_{I'}^{k})
(\mathcal{U}_{I}^{\bar j}\partial_{\bar j}\mathcal{U}_{I'}^{\bar l}) -
(\mathcal{U}_{I'}^{i}\partial_{i}\mathcal{U}_{I}^{k})
(\mathcal{U}_{I}^{\bar j}\partial_{\bar j}\mathcal{U}_{I'}^{\bar l})
\right),
\end{eqnarray}
where $\mathcal{U}_{I}^{i}=\mathcal{U}_{I}^{i}(X)$ and $\mathcal{U}_{I}^{\bar i}=
\mathcal{U}_{I}^{\bar i}(\bar X)$ are holomorphic and anti-holomorphic
"blocks" for the background metric field. Multiplying \rf{cond1alg} from the right by
$\partial_k\partial_{\bar l}$, one can rewrite \rf{cond1alg} as
\begin{eqnarray}\label{dcom}
&&\left(g^{i\bar{j}}\partial_{i}\partial_{\bar{j}}g^{k\bar{l}}-
\partial_{i}g^{k\bar{j}}\partial_{\bar j}g^{i\bar{l}}\right)\partial_k\partial_{\bar l}=
\nonumber
\\
&&=\sum_{I,I'}[{v}_{I}, v_{I'}]{\bar v}_I{\bar v}_{I'} =
{1\over 2}\sum_{I,I'}[{v}_{I}, v_{I'}][{\bar v}_I,{\bar v}_{I'}] = 0,
\end{eqnarray}
where we have introduced vector fields
$v_I=\mathcal{U}_{I}^{i}\partial_i$ and ${\bar v}_I=
\mathcal{U}_{I}^{\bar i}\partial_{\bar i}$.
For the r.h.s. of \rf{dcom}, it is convenient to use the notation
\begin{eqnarray}\label{dcomm}
&&[[g,\tilde g]](X,p,\bar{X},\bar{p})=\sum_{I,J}
[\mathcal{U}_{I},\tilde\mathcal{U}_{J}](X,p)\otimes[\bar{\mathcal{U}}_{I},
\bar{\tilde{\mathcal{U}}}_{J}](\bar{X},\bar{p}),
\end{eqnarray}
where $g=g^{i\bar{j}}{\p_ i\p_{\bar{j}}}$ is a bivector field. 
Now, we are ready to express the structure of equations \rf{feinst}-\rf{inhom} in an invariant way.\\
{\bf Proposition 4.3.} 
1). {\it Equation \rf{vol} means that it is possible to decompose $\Phi_0$ on holomorphic and antiholomorphic 
part, that is, one can consider the following object: 
\begin{eqnarray}
\Omega(X)\b \Omega(\b X)=e^{-(2\Phi_0)},  
\end{eqnarray}
where $\Omega(X) dX^1\wedge...\wedge dX^n$ ($\b \Omega(\b X) dX^{\b 1}\wedge...\wedge dX^{\b n}$) is a 
holomorphic (antiholomorphic) volume form.} \\
2). {\it Equation \rf{hom} means that the 
appropriate parts of the vector field 
\begin{eqnarray}
div_{\Omega}(g)=\Omega^{-1}\p_i(\Omega g^{i\b j})\p_{\b j}+
\b \Omega^{-1}\p_{\b i}(\b \Omega g^{\b ij})\p_j
\end{eqnarray}
belong to respectively holomorphic and antiholomorphic vector bundles.} \\
3). {\it Equation \rf{inhom} can be rewritten in the following way:
\begin{eqnarray}
[[g,g]]+\mathcal{L}_{div_{\Omega}(g)}g=0, \quad 
\end{eqnarray}
where $[[,]]$ is a double commutator for the bivector field $g^{i\b j}\p_i\p_{\b j}$ 
and $\mathcal{L}_{div_{\Omega}(g)}$ is a Lie derivative with respect to the 
vector field $div_{\Omega}(g)$}.
\section{The Nonlinear Sigma Model}
Let's consider the theory of D free massless bosons with the action: 
\begin{eqnarray} 
S=\frac{1}{4\pi\alpha'}\int d^2 z \eta_{\mu\nu}\p X^{\mu}\bp X^{\nu},
\end{eqnarray}
where  $\eta_{\mu\nu}$ is a constant nondegenerate symmetric matrix, 
$\mu,\nu=1,...,D$ and $d^2z=idz\wedge d\b z$.\\
\hspace*{5mm}The operator product, generated by the free boson field theory, is as follows:
\begin{eqnarray}
X^{\alpha}(z_1)X^{\beta}(z_2)\sim -\eta^{\alpha\beta}\alpha'\log|z_{12}/\mu|^2,
\end{eqnarray}
where $\mu$ is some nonzero parameter. 
The energy-momentum tensor is given by such an expression: 
\begin{eqnarray}
T=-(2\alpha')^{-1}\p X^{\mu}\p X_{\mu}, \quad 
\t T=-(2\alpha')^{-1}\bp X^{\mu}\bp X_{\mu}.
\end{eqnarray}\\
Let's consider the usual nonlinear sigma-model action
\begin{eqnarray} \label{sigma}
S=\frac{1}{4\pi\alpha'}\int d^2 z G_{\mu\nu}(X)\p X^{\mu}\bp X^{\nu},
\end{eqnarray}
where $G_{\mu\nu}(X)$ is expanded with respect to some formal parameter:
\begin{eqnarray}\label{gexp} 
G_{\mu\nu}=\eta_{\mu\nu}-th_{\mu\nu}(X)-t^2s_{\mu\nu}(X)+O(t^3).
\end{eqnarray}
This allows dealing with 
$\phi^{(2)}=(2\alpha')^{-1}(\eta_{\mu\nu}-G_{\mu\nu})\p X^{\mu}\bp X^{\nu}dz\wedge d\b z$ as 
a perturbation 2-form and applying our Master Equation to this case. 
But it appears that in such a way we can miss something important. 
Firstly, let's consider symmetries. 
Action \rf{sigma} is invariant under the diffeomorphism transformations, under which 
the infinitesimal change of metric $G_{\mu\nu}$ is given by:
\begin{eqnarray}\label{gt}
G_{\mu\nu}\to G_{\mu\nu}-\nabla_{\mu}v_{\nu}-\nabla_{\nu}v_{\mu},
\end{eqnarray}   
where $v_{\nu}$ are infinitesimal. Let's expand
\begin{eqnarray}\label{vexp}
v_{\nu}=tv^1_{\nu}+t^2v^2_{\nu}+O(t^3).
\end{eqnarray}   
Let's consider the following operator-valued 2-forms:
\begin{eqnarray}
&&\phi_{G,1}^{(2)}=dz\wedge d\b z(2\alpha')^{-1}h_{\mu\nu}(X)\p X^{\mu}\bp X^{\nu},\\
&&\phi_{G,2}^{(2)}=dz\wedge d\b z((2\alpha')^{-1}(\t s_{\mu\nu}(X)\p X^{\mu}\bp X^{\nu}+\epsilon-terms),\nonumber
\end{eqnarray} 
where $h_{\mu\nu}$ and $\t s_{\mu\nu}$ are symmetric. According to \rf{symm}, the following transformations 
\begin{eqnarray}\label{sy}
&&\delta \phi_{G,1}^{(2)}(z)=-d\xi_1^{(1)}(z),\\
&&\delta \phi_{G,2}^{(2)}(z)=-d(\xi_2^{(1)}(z)+\chi^{(1)}(\xi_1^{(1)},\phi_1^{(1)}) ) +
\frac{1}{2\pi i}\int_{C_{\epsilon, z}}\xi_1^{(1)}(w)\phi_1^{(2)}(z)\nonumber
\end{eqnarray}
give the symmetries of the Master equation. We put 
 \begin{eqnarray}
\xi_i^{(1)}=(2\alpha')^{-1}(v^i_{\mu}(X)\p X^{\mu}dz - v^i_{\mu}(X)\b \p X^{\mu}d\b z)\quad (i=1,2).
\end{eqnarray} 
Choosing  
\begin{eqnarray}
\chi^{(1)}(\xi_1^{(1)},\phi_{1,G}^{(2)})=(2\alpha')^{-1}v^1_{\rho}\eta^{\rho\mu}h_{\mu\nu}(\p X^{\mu}dz-\bp X^{\mu}d\b z),
\end{eqnarray}
one finds that \rf{sy} is equivalent to
\begin{eqnarray}\label{sym2}
&&\delta \phi_{G,1}^{(2)}(z)=(2\alpha')^{-1}
(\p_{\mu}v^1_{\nu}+\p_{\nu}v^1_{\mu})\p X^{\mu}\bp X^{\nu}dz\wedge d\b z,\\
&&\delta \phi_{G,2}^{(2)}(z)=(2\alpha')^{-1}
(\p_{\mu}(v^2_{\nu}+v^1_{\rho}\eta^{\rho\lambda}h_{\lambda\nu})+\nonumber\\
&&\p_{\nu}(v^2_{\mu}+
v^1_{\rho}\eta^{\rho\lambda}h_{\lambda\mu}))\p X^{\mu}\bp X^{\nu}dz\wedge d\b z+\nonumber\\
&&\frac{1}{2\pi i}\int_{C_{\epsilon, z}}(1/2\alpha'^{-1}v^1_{\mu}\p X^{\mu}dz'-
1/2\alpha'^{-1}v^1_{\mu}\bp X^{\mu}d\b z')\nonumber\\
&&(2\alpha')^{-1}h_{\mu\nu}\p X^{\mu}\bp X^{\nu}(z)dz\wedge d\b z
\end{eqnarray} 
and leads to the following transformations of fields $h_{\mu\nu}$ and $\t s_{\mu\nu}$:
\begin{eqnarray}
&&\delta h_{\mu\nu}=\p_{\mu}v^1_{\nu}+\p_{\nu}v^1_{\mu},\\
&&\delta \t s_{\mu\nu}=\delta s_{\mu\nu}+
1/2\delta (h_{\mu\rho}\eta^{\rho\sigma}h_{\nu\sigma})+O(\alpha'),
\end{eqnarray} 
where 
\begin{eqnarray}
&&\delta s_{\mu\nu}=\p_{\mu}v^2_{\nu}+\p_{\nu}v^2_{\mu}-2\Gamma^{\rho}_{\mu\nu}v^1_{\rho}=\nonumber\\
&&\p_{\mu}v^2_{\nu}+\p_{\nu}v^2_{\mu}+v^1_{\rho}\eta^{\rho\sigma}(-\p_{\sigma}h_{\mu\nu}+
\p_{\mu}h_{\sigma\nu}+\p_{\nu}h_{\sigma\mu})
\end{eqnarray} 
is the standard diffeomorphism change at the second order of expansion of $G_{\mu\nu}$ \rf{gexp}. 
Then, it is reasonable to put
\begin{eqnarray}
\t s_{\mu\nu}=s_{\mu\nu}+1/2h_{\mu\rho}\eta^{\rho\sigma}h_{\nu\sigma}
\end{eqnarray} 
and thus to consider such perturbation operators
\begin{eqnarray}
&&\phi_{G,1}^{(2)}=dz\wedge d\b z(2\alpha')^{-1}h_{\mu\nu}(X)\p X^{\mu}\bp X^{\nu},\\
&&\phi_{G,2}^{(2)}=dz\wedge d\b z((2\alpha')^{-1}( s_{\mu\nu}(X)\p X^{\mu}\bp X^{\nu}+\nonumber\\
&&1/2h_{\mu\rho}(X)\eta^{\rho\sigma}h_{\nu\sigma}(X)\p X^{\mu}\bp X^{\nu}+\epsilon-terms).
\end{eqnarray}
The second term in $\phi_{G,2}^{(2)}$ seems to be quite strange because at the first glance this term is absent in the 
expansion of the action of the nonlinear sigma-model.\\ 
\hspace*{5mm}The necessity of such term can be substantiated 
in the following way. Our method of regularization is very natural on the operator level but, 
definitely, we lose some terms in comparison with the usual Feynman diagram approach. 
Such terms arise when there is a Wick contraction leading to delta function or its derivatives, 
giving local terms in the action after integration, say for example 
\begin{eqnarray}
\int d^2z_1 \alpha'^{-1}h_{\mu\nu}\p X^{\mu}\bp X^{\nu}(z_1)
\int d^2z_2 \alpha'^{-1}h_{\rho\sigma}\p X^{\rho}\bp X^{\sigma}(z_2)
\to \nonumber\\
\int d^2z_1\int d^2z_2 4\alpha'^{-1}\pi\delta^{(2)}(z_1-z_2)h_{\mu\rho}\eta^{\rho\sigma}h_{\nu\sigma}\p X^{\mu}\bp X^{\nu}(z_2).
\end{eqnarray} 
The only way to include them in the game is simply to add them to the perturbation as we did. 
The next proposition assures that we made the right choice for the perturbation operators.\\
{\bf Proposition 5.1.} 
{\it Constraints \rf{m1} and \rf{m2} for 
 \begin{eqnarray}
&&V_1=(2\alpha')^{-1}h_{\mu\nu}(X)\p X^{\mu}\bp X^{\nu},\nonumber\\
&&V_2=(2\alpha')^{-1}(s_{\mu\nu}(X)\p X^{\mu}\bp X^{\nu}+
1/2h_{\mu\rho}(X)\eta^{\rho\sigma}h_{\nu\sigma}(X)\p X^{\mu}\bp X^{\nu}),
\end{eqnarray}
$U_i\equiv f_i(X)$ and $\b U_i\equiv \b f_i(X)$ in the order ${\alpha'}^0$ give the 
Einstein equations up to the second order in t: 
\begin{eqnarray}
R_{\mu\nu}+2\nabla_{\mu}\nabla_{\nu}\Phi=0
\end{eqnarray}
with a metric 
\begin{eqnarray} 
G_{\mu\nu}=\eta_{\mu\nu}-th_{\mu\nu}(X)-t^2s_{\mu\nu}(X)+O(t^3),
\end{eqnarray}
and a dilaton
\begin{eqnarray}
&&\Phi=\Phi_0+1/2t(f_1+\b f_1-1/2h)+\nonumber\\
&&1/2t^2(f_2+\b f_2-1/2s-1/8h_{\mu\nu}h^{\mu\nu})+O(t^3),
\end{eqnarray}
where $h=\eta^{\mu\nu}h_{\mu\nu}$ and $s=\eta^{\mu\nu}s_{\mu\nu}$, also 
$\eta_{\mu\nu}$ and $\Phi_0$ are independent of $X$.}\\
The proof is given in Appendix C.
\section{Conclusion and Outlook}
In this paper, we have started our study of the operator equations of conformal invariance 
from the equation expressing 
the conservation law of the current in the presence of a perturbation (represented in the Maurer-Cartan form): 
\begin{eqnarray}\label{cons}
D\Phi(z,\theta)+\int_{C_{\epsilon, z}}\Phi(w,\theta)\Phi(z,\theta)+...=0,
\end{eqnarray}
where $D=d+\theta Q_B$, $\Phi=\phi^{(2)}+\theta\psi^{(1)}$, and $\theta$ is some Grassmann variable, anticommuting with $d$.  
Studying this equation up to the second order in the formal parameter 
$\Phi=\sum^{\infty}_{i=1}t^i\Phi_i$, we learned that there are some ambiguities in the solution leading 
to the appearance of holomorphic and antiholomorphic operators without a clear physical explanation. 
But at the same time, we have found a way, how one can get rid of these operators, 
reducing a number of independent 
variables from two ($\phi^{(2)}, \psi^{(1)}$) to one $\phi^{(0)}$ 
such that equation \rf{cons} is equivalent to the Master Equation: 
\begin{eqnarray}\label{mast}
P([Q_B,\phi^{(0)}]+\frac{1}{4\pi i}\int_{C_{\epsilon,z}} \phi^{(1)}(w)\phi^{(0)}(z)+...)=0,
\end{eqnarray}
where $P=b_{-1}\t b_{-1}$ and
$
\phi^{(1)}=dz[b_{-1},\phi^{(0)}]+d\b z[\t b_{-1},\phi^{(0)}].
$ 
Here dots mean higher operations in 
$\phi^{(0)}$.\\ 
\hspace *{5mm} Since the expression 
$[Q_B,\phi^{(0)}]+\frac{1}{4\pi i}\int_{C_{\epsilon,z}} \phi^{(1)}(w)\phi^{(0)}(z)$ is of ghost number 3 and 
we expect that 
both $\phi^{(1)}$ and $\phi^{(0)}$ are present in the higher operations, the reasonable expression for 
the higher operation is: 
\begin{eqnarray}\label{hop}
\int \phi^{(1)}(w)\phi^{(0)}(z)\int \phi^{(2)}(z_1)...\int \phi^{(2)}(z_n).
\end{eqnarray} 
The problem is to find the region of integration in the formula \rf{hop} and the proper coefficients 
$c_n$ with which expression \rf{hop} enters \rf{mast}. \\
\hspace*{5mm}
The next problem is to find the symmetry transformations of the given equations. This will allow 
expressing the complicated 
target-space symmetries of string theory in the background fields (e.g. diffeomorphisms, 
etc) on the operator language.\\  
\hspace*{5mm} Another topic, which we did not touch in this paper is the study of perturbations of CFT with 
boundary conditions, 
e.g. analogue of equations \rf{cons} and \rf{mast} in the case of open strings in the background fields. 
For example, it is well-known that the Maxwell equations for a background abelian gauge field 
is (see e.g. \cite{pol}):
\begin{eqnarray}\label{ym}
[Q_B, \phi^{(0)}_A]=0,
\end{eqnarray}
where $\phi^{(0)}_A=cA_{\mu}\p X^{\mu}-1/2\p c \p^{\mu} A_{\mu}$. For nonabelian gauge field, the equations should 
include more terms and to obtain corresponding Yang-Mills equations one should consider higher order corrections 
to equation \rf{ym}. We will consider the Master equation in the case of open strings in a separate 
publication.

\section*{Acknowledgements}
The author is very grateful to A. Losev and A. Marshakov for 
introduction in the subject and many fruitful discussions and remarks. 
It is important to mention that the proposal concerning the using of generalized Maurer-Cartan equations 
in the context of study of conditions of conformal invariance belongs to A. Losev.
The author is very grateful to A.A. Tseytlin for his comments.\\ \hspace*{3mm}
The research is supported by the Dynasty Foundation, CRDF (Grant No. RUM1-2622-ST-04) 
and RFBR (05-01-00922).

\section*{Appendix A: Proof of Proposition 3.2.}
The equation
\begin{eqnarray}\label{mea1}
[Q_B,\phi_1^{(2)}]=d\psi_1^{(1)},
\end{eqnarray}
where $\phi_1^{(2)}=V_1(z)dz\wedge d\b z$ ($V_1$ is of type $(1,1)$) and 
\begin{eqnarray}
&&\psi_1^{(1)}=(c (Y_1-(\p c+\bp \tc)\b W_1 -1/2\bp^2 \tc \b U_1+\tc \b Z_1)d\b z
+\nonumber\\
&&(-\tc \b Y_1+(\p c+\bp \tc)W_1+1/2\p^2 c U_1-cZ_1)dz,
\end{eqnarray}
leads to the following relations between the components:
\begin{eqnarray}\label{fm10}
&&L_{-1}V_1=L_{-1}Y_1+\b L_{-1}Z_1, \nonumber\\
&&\b L_{-1}V_1=\b L_{-1}\b Y_1+ L_{-1}\b Z_1,\nonumber\\
&&(L_0V_1)-Y_1+L_{-1}\b W_1+\b L_{-1} W_1=0,\nonumber\\
&&(L_0V_1)-\b Y_1+L_{-1}\b W_1+\b L_{-1} W_1=0,\nonumber\\
&&L_1V_1+2\b W_1+\b L_{-1} U_1=0,\nonumber\\
&&\b L_1V_1+2W_1+L_{-1} \b U_1=0.
\end{eqnarray}
The third and the fourth relations give $Y_1=\b Y_1$ and thus the first and the second ones can be  
represented as 
\begin{eqnarray}
L_{-1}(V_1-Y_1)=\b L_{-1}Z_1\nonumber\\
\b L_{-1}(V_1-\b Y_1)=L_{-1}\b Z_1.
\end{eqnarray}
Hence, $\b L_{-1}\b L_{-1}Z_1=L_{-1}L_{-1}\b Z_1$. This relation in its turn gives the following 
representation for $Z_1$ and $\b Z_1$: 
\begin{eqnarray}
&&Z_1=Z^h_1+L_{-1}L_{-1}R_1,\nonumber\\
&&\b Z_1=\b Z^a_1+\b L_{-1}\b L_{-1}R_1,
\end{eqnarray}
where $R_1$ is some operator, $Z^h_1$ is a holomorphic operator, and $\b Z^a_1$ is an 
antiholomorphic one. Therefore, $Y_1=\b Y_1=V_1-L_{-1}\b L_{-1}R_1$. 
Just for the case, to be sure that all is correct, we give the solutions of the 
equation $d{\psi'}_1^{(1)}=0$ also. 
The solution of this equation will give the ambiguity of the solution of equation \rf{mea1}.
To find the solution one should just put 
$V_1$ to zero in the equations \rf{fm10} and thus lead to very simple relations:
\begin{eqnarray} 
&&Y'_1=\b Y'_1=-1/2L_{-1}\b L_{-1}(U'_1+\b U'_1),\nonumber\\
&&W'_1=-1/2L_{-1}\b U'_1,\quad \b W'_1=-1/2\b L_{-1}U'_1,\nonumber\\
&&L_{-1}Y'_1+\b L_{-1}Z'=0,\quad \b L_{-1}\b Y'_1+L_{-1}\b Z'=0,
\end{eqnarray}
where we used prime to differ the solutions of these equations from the Master equation.
It is easy to see that the last pair of equations leads to 
\begin{eqnarray} 
&&Z'_1=Z'^h_1+1/2L_{-1}L_{-1}(U'_1+\b U'_1), \nonumber\\ 
&&\b Z'_1=\b Z'^a_1+1/2\b L_{-1}\b L_{-1}(U'_1+\b U'_1).
\end{eqnarray}
Moreover, this leads to the fact that all solutions of such an equations have the form
\begin{eqnarray} 
{\psi'}_1^{(1)}=d{\psi'}_1^{(0)}-J_{Z'_1},
\end{eqnarray}
where 
\begin{eqnarray} 
&&{\psi'}_1^{(0)}=-1/2c\p (U'_1+\b U'_1)+1/2\p c U'_1+1/2\t c\bp (U'_1+\b U'_1)-1/2\bp\t c \b U'_1,\nonumber\\
&&J_{Z'_1}=cZ'^h_1dz-\t c Z'^a_1d\b z.
\end{eqnarray}
The first term just corresponds to the obvious symmetry of the deformed charge, the other one gives the ambiguity 
of the solution of the master equation at the first order in $t$.\\ 
So, the complete solution of \rf{me1} is of the following kind: 
\begin{eqnarray} 
&&\psi_{1,J}^{(1)}=(cV_1 -(\p c+\bp \tc)\b W_1 -1/2\bp^2 \tc \b U_1)d\b z
+\nonumber\\
&&(-\tc \b V_1+(\p c+\bp \tc)W_1+1/2\p^2 c U_1)dz +J_{Z_1},
\end{eqnarray}
where according to system \rf{fm11} we have the following relations (absorbing $R_1$ inside $U_1$ and $\b U_1$ by 
means of addition of the total derivative):
\begin{eqnarray}\label{fm11}
&&(L_0V_1)(z)-V_1(z)-1/2L_{-1}L_1V_1-1/2\b L_{-1}\b L_1V_1-\\
&&1/2L_{-1}\b L_{-1} (U_1+\b U_1)=0,\nonumber\\
&&\b W_1=-1/2((L_1V)(z)+\b L_{-1}U_1),\quad W_1=-1/2((\b L_1V)(z)+L_{-1}\b U_1).\nonumber
\end{eqnarray}
Studying the second order, we put $J_{Z_1}=0$ 
according to point 2 of the Proposition 3.2. 
So, we consider the second equation:
\begin{eqnarray}
[Q,\phi_2^{(2)}](z)+\frac{1}{2\pi i}\int_{C_{\epsilon, z}}\psi^{(1)}_1\phi_1^{(2)}(z)=d\psi_2^{(1)}(z) 
\end{eqnarray}
with $\psi^{(1)}_1\equiv \psi^{(1)}_{1,0}$.  
Here, we have the quadratic operation between $\psi_{1}^{(1)}$ and 
$\phi_{1}^{(2)}$:  
Really, 
\begin{eqnarray}
&&\int_{C_{\epsilon,z}}\psi_{1}^{(1)}(w)
\phi_{1}^{(2)}(z)=(-\p c (V_1,V_1)_0^{(1,1)}-\bp \t c (V_1,V_1)_0^{(1,1)}-\nonumber\\
&&1/2\p^2c(V_1,V_1)_0^{(2,1)}-1/2\bp^2\t c(V_1,V_1)_0^{(1,2)}-c(V_1,V_1)_0^{(0,1)}-\nonumber\\
&&\tc (V_1,V_1)_0^{(1,0)}+
(\p c+\bp\tc)(\b W_1,V_1)_0^{(0,1)} +(\p c+\bp\tc)(W_1,V_1)_0^{(1,0)}+\nonumber\\
&&1/2\bp^2\tc(\b U_1,V_1)_0^{(0,1)}+1/2\p^2 c(\b U_1,V_1)_0^{(1,0)}+\bp^2\tc((\b W_1,V_1)_0^{(0,2)}+\nonumber\\
&&(W_1,V_1)_0^{(1,1)})+1/2\p^2 c((W_1,V_1)_0^{(2,0)}+(\b W_1,V_1)_0^{(1,1)})+\nonumber\\
&&\epsilon-terms)dz\wedge d\b z=((\p c+\bp c)(-1/2(V_1,V_1)_0^{(1,1)}+(\b W_1,V_1)_0^{(0,1)}+\nonumber\\
&&(W_1,V_1)_0^{(1,0)})+\p^2c((W_1,V_1)_0^{(2,0)}+(\b W_1,V_1)_0^{(1,1)}+
1/2(U_1,V_1)_0^{(1,0)})+\nonumber\\
&&\bp^2\t c((\b W_1,V_1)_0^{(0,2)}+(W_1,V_1)_0^{(1,1)}+1/2(\b U_1,V_1)_0^{(0,1)})+\nonumber\\
&&\bp K_2+ \p \b K_2+\epsilon-terms)dz\wedge d\b z,
\end{eqnarray}
where 
\begin{eqnarray}
&&K_2=-1/4\p^2c(V_1,V_1)_0^{(2,2)}+1/4c\p^2(V_1,V_1)_0^{(2,2)}-\nonumber\\
&&1/2\t c (V_1,V_1)_0^{(1,1)}-1/2c (V_1,V_1)_0^{(0,2)}\nonumber\\
&&\b K_2= -1/4\bp^2\t c(V_1,V_1)_0^{(2,2)}+1/4\tc\bp^2(V_1,V_1)_0^{(2,2)}-\nonumber\\
&&1/2c (V_1,V_1)_0^{(1,1)}-1/2\tc (V_1,V_1)_0^{(2,0)}.
\end{eqnarray}
This motivates to express $\psi_{2}^{(1)}$ in the following way:
\begin{eqnarray}
&&\psi_2^{(1)}=(c Y_2 -(\p c+\bp \tc)\b W_2 -1/2\bp^2 \tc \b U_2+\tc \b Z_2+\b K_2)d\b z
+\nonumber\\
&&(-\tc \b Y_2+(\p c+\bp \tc)W_2+1/2\p^2 c U_2-cZ_2-K_2)dz+\epsilon-terms.
\end{eqnarray}
This leads to the following system of equations:
\begin{eqnarray}
&&L_{-1}V_2=L_{-1}Y_2+\b L_{-1}Z_2, \\
&&\b L_{-1}V_2=\b L_{-1}\b Y_1+ L_{-1}\b Z_2, \nonumber\\
&&(L_0V_2)(z)-Y_2(z)-1/2(V_1,V_1)_0^{(1,1)}(z)+(\b W_1,V_1)_0^{(0,1)}(z)+\nonumber\\ 
&&(W_1,V_1)_0^{(1,0)}(z)+L_{-1}\b W_2+ \b L_{-1} W_2=0,\nonumber\\
&&(L_0V_2)(z)-\b Y_2(z)-1/2(V_1,V_1)_0^{(1,1)}(z)+(\b W_1,V_1)_0^{(0,1)}(z)+\nonumber\\ 
&&(W_1,V_1)_0^{(1,0)}(z)+L_{-1}\b W_2+ \b L_{-1} W_2=0,\nonumber\\
&&(\b W_1,V_1)^{(1,2)}(z)+
(W_1,V_1)_0^{(2,1)}(z)-1/2(V_1,V_1)_0^{(2,2)}(z)=0\nonumber\\
&&(L_1 V_2)+2\b W_2+2(\b W_1,V_1)_0^{(1,1)}(z)+2(W_1,V_1)_0^{(2,0)}(z)\nonumber\\
&&+(U_1,V_1)_0^{(1,0)}+ \b L_{-1}U_2=0,\nonumber\\
&&(\b L_1 V_2)+2W_2+2(W_1,V_1)_0^{(1,1)}(z)+2(\b W_1,V_1)_0^{(0,2)}(z)\nonumber\\
&&+(\b U_1,V_1)_0^{(0,1)} + L_{-1}\b U_2=0\nonumber
\end{eqnarray}
The first two of them are familiar to us from the first order case. As in that situation, the solution is given 
by the following relations:
\begin{eqnarray}
&&Z_2=Z^h_2+L_{-1}L_{-1}R_2,\nonumber\\
&&\b Z_2=\b Z^a_2+\b L_{-1}\b L_{-1}R_2,\nonumber\\
&&Y_2=\b Y_2=V_2-L_{-1}\b L_{-1}R_2,
\end{eqnarray}
where  $Z^h_2$ is a holomorphic operator, $\b Z^a_2$ is an antiholomorphic one, and $R_2$ is some operator. 
Again, as in first order case, one can absorb the $R_2$ terms in $U_2$ and $\b U_2$ by adding the total derivative
to $\psi_{2}^{(1)}$, so the complete system of equations at the second order is:
\begin{eqnarray}
&&(L_0V_2)(z)-V_2(z)-1/2(V_1,V_1)_0^{(1,1)}(z)-1/2
( L_1V_1+\b L_{-1} U_1,V_1)_0^{(0,1)}(z)-\nonumber\\ 
&& 1/2(\b L_1V_1+L_{-1} \b U_1,V_1)_0^{(1,0)}(z)+L_{-1}\b W_2(z)+ \b L_{-1} W_2(z)=0,\nonumber\\
&&\b W_2(z)=-1/2((L_1 V_2)(z)-(L_1V_1+\b L_{-1} U_1,V_1)_0^{(1,1)}(z)-\nonumber\\
&&(\b L_1V_1+L_{-1} \b U_1,V_1)_0^{(2,0)}(z)+(U_1,V_1)_0^{(1,0)}(z)+\b L_{-1}U_2(z))\nonumber\\
&&W_2(z)=-1/2((\b L_1 V_2)(z)-(\b L_1V_1+L_{-1} \b U_1,V_1)_0^{(1,1)}(z)-\nonumber\\
&&(L_1V_1+\b L_{-1} U_1,V_1)_0^{(0,2)}(z)+(\b U_1,V_1)_0^{(0,1)}(z)+ L_{-1}\b U_2(z)).
\end{eqnarray}

\addcontentsline{toc}{section}{Appendix B: Proof of Proposition 4.2.}
\section*{Appendix B: Proof of Proposition 4.2.}
We use the following formula for the Ricci tensor \cite{fock}:
\begin{eqnarray}
R^{\mu\nu}= 
1/2  G^{\alpha\beta}\partial_{\alpha}\partial_{\beta}G^{\mu\nu}+
\Gamma^{\mu\nu}-\Gamma^{\mu,\alpha\beta}\Gamma^{\nu}_{\alpha\beta},
\end{eqnarray}
where 
$$
\Gamma^{\mu\nu}=G^{\mu\rho}G^{\nu\sigma}\Gamma_{\rho\sigma},\quad
\Gamma_{\rho\sigma}=\half  (\partial_{\rho}\Gamma_{\sigma}+\partial_{\sigma}
\Gamma_{\rho})-\Gamma^{\nu}_{\rho\sigma}\Gamma_{\nu},
$$
$$
\Gamma_{\nu}=G^{\alpha\beta}\partial_{\beta}G_{\alpha\nu}-
\half \partial_{\nu}\log (G).
$$
Remembering that $G_{i\bar{k}}=g_{i\bar{k}}$ and $B_{i\bar{k}}=-g_{i\bar{k}}$, this leads to: 
\begin{eqnarray}\label{ricci}
&&R^{\mu\nu}-{1\over 4} H^{\mu\lambda\rho}H^{\nu}_{\lambda\rho}+2\nabla^{\mu}\nabla^{\nu}\Phi= \nonumber\\
&&2\nabla^{\mu}\nabla^{\nu}\Phi_0-{1\over 4} H^{\mu\lambda\rho}H^{\nu}_{\lambda\rho}+
1/2  G^{\alpha\beta}\partial_{\alpha}\partial_{\beta}G^{\mu\nu}+
\Gamma'^{\mu\nu}-\Gamma^{\mu,\alpha\beta}\Gamma^{\nu}_{\alpha\beta},
\end{eqnarray}
where 
$$
\Gamma'_{\rho\sigma}=\half  (\partial_{\rho}\Gamma'_{\sigma}+\partial_{\sigma}
\Gamma'_{\rho})-\Gamma^{\nu}_{\rho\sigma}\Gamma'_{\nu},
$$
$$ 
\Gamma'_{\nu}=G^{\alpha\beta}\partial_{\beta}G_{\alpha\nu}, \quad \Phi_0=\Phi-\log \sqrt{g},
$$
and $g$ is the determinant of matrix $g_{i\bar{j}}$.
Now let us study the third term in (\ref{ricci}): first,
for the components of $ \Gamma^{\nu}_{\alpha\beta}$, one has:
\begin{eqnarray}\label{crist}
&&\Gamma^{i}_{rs}=\half  g^{i\bar{k}}(\partial_{r}g_{\bar{k}s}+
\partial_{s}g_{\bar{k}r}),  \nonumber\\
&&\Gamma^{i}_{r\bar{s}}=\half g^{i\bar{k}}(\partial_{\bar{s}}g_{r\bar{k}}
-\partial_{\bar{k}}
g_{r\bar{s}}) \quad and \quad c.c.,
\end{eqnarray}
while all other components vanish. Therefore, one finds that
$\Gamma_{\bar{i},r\bar{s}}=\half H_{\bar{s}\bar{i}r}$, hence the third term in
(\ref{ricci}) provides the contribution of the $H^2$-type with an additional
term in $\Gamma\Gamma$ for $\mu=\bar{i}$ and $\nu=j$:
\begin{eqnarray}
&&\Gamma^{\bar{i},kl}\Gamma^{j}_{kl}=-{1\over 4} (g^{k\bar{r}}\partial_{\bar{r}}
g^{l\bar{i}}+g^{l\bar{r}}\partial_{\bar{r}}
g^{k\bar{i}})g^{j\bar{p}}(\partial_{k}g_{\bar{p}l}+\partial_{l}g_{\bar{p}k})=
\nonumber\\
&&-{1\over 4} (g^{k\bar{r}}\partial_{\bar{r}}
g^{l\bar{i}}-g^{l\bar{r}}\partial_{\bar{r}}
g^{k\bar{i}})g^{j\bar{p}}(\partial_{k}g_{\bar{p}l}-\partial_{l}g_{\bar{p}k})-
g^{k\bar{r}}\partial_{\bar{r}}
g^{l\bar{i}}g^{j\bar{p}}\partial_{l}g_{\bar{p}k}=\nonumber\\
&&-{1\over 4}H^{\bar{i}kl}H^{j}_{kl}+\partial_{\bar{r}}g^{\bar{i}k}
\partial_{k}g^{\bar{r}j}.
\end{eqnarray}
Thus we can see that
\begin{eqnarray}
&&R^{i\bar{k}}-{1\over 4} H^{i\lambda\rho}H^{\bar{k}}_{\lambda\rho}+
2\nabla^{i}\nabla^{\bar{k}}\Phi=\nonumber\\
&&2g^{r\bar{l}}\p_r\p_{\bar{l}}g^{i\bar{k}}-2\p_r g^{i\bar{p}}\p_{\bar{p}}g^{r\bar{k}}-
(\nabla^i\xi^{\bar{k}}+
\nabla^{\bar{k}}\xi^i)=\nonumber\\
&&2g^{r\bar{l}}\p_r\p_{\bar{l}}g^{i\bar{k}}-2\p_r g^{i\bar{p}}\p_{\bar{p}}g^{r\bar{k}}-
g^{i\bar{l}}\p_{\bar{l}}d^{\Phi_0}_sg^{s\bar{k}}-g^{r\bar{k}}\p_r d^{\Phi_0}_{\bar{j}}g^{\bar{j}i}
+\nonumber\\
&&\p_rg^{i\bar{k}}d^{\Phi_0}_{\bar{j}}g^{\bar{j}r}+\p_{\bar{p}}g^{\bar{k}i}d^{\Phi_0}_n g^{n\bar{p}},
\end{eqnarray}
where $\xi^{\mu}=d^{\Phi_0}_{\nu}G^{\nu\mu}=\partial_{\nu}G^{\nu\mu}-2\p_{\nu}\Phi_0G^{\nu\mu}$ and 
\begin{eqnarray}
&&R^{ij}-{1\over 4} H^{i\lambda\rho}H^{j}_{\lambda\rho}+
2\nabla^{i}\nabla^{j}\Phi=\nonumber\\
&&\nabla^i\xi^{j}+\nabla^{j}\xi^i=g^{i\bar{p}}\p_{\bar{p}}d^{\Phi_0}_{\bar{l}}g^{\bar{l}k}+
g^{k\bar{p}}\p_{\bar{p}}d^{\Phi_0}_{\bar{l}}g^{\bar{l}i}.
\end{eqnarray}
Hence the equation 
$R^{\mu\nu}-{1\over 4} H^{\mu\lambda\rho}H^{\nu}_{\lambda\rho}+2\nabla^{\mu}\nabla^{\nu}\Phi=0$
leads to the equations:
\begin{eqnarray}
&&2g^{r\bar{l}}\p_r\p_{\bar{l}}g^{i\bar{k}}-2\p_r g^{i\bar{p}}\p_{\bar{p}}g^{r\bar{k}}-
g^{i\bar{l}}\p_{\bar{l}}d^{\Phi_0}_sg^{s\bar{k}}-g^{r\bar{k}}\p_r d^{\Phi_0}_{\bar{j}}g^{\bar{j}i}
+\nonumber\\
&&\p_rg^{i\bar{k}}d^{\Phi_0}_{\bar{j}}g^{\bar{j}r}+
\p_{\bar{p}}g^{\bar{k}i}d^{\Phi_0}_n g^{n\bar{p}}=0,\nonumber\\
&&g^{i\bar{p}}\p_{\bar{p}}d^{\Phi_0}_{\bar{l}}g^{\bar{l}k}+
g^{k\bar{p}}\p_{\bar{p}}d^{\Phi_0}_{\bar{l}}g^{\bar{l}i}=0\quad and \quad c.c..
\end{eqnarray}
Similarly one can show that the equations 
\begin{eqnarray}
\nabla_{\mu}H^{\mu\nu\rho}-2(\nabla_{\lambda}\Phi)H^{\lambda\nu\rho}=0
\end{eqnarray}
leads to an additional system of equations:
\begin{eqnarray}
&&g^{i\bar{p}}\p_{\bar{p}}d^{\Phi_0}_{\bar{l}}g^{\bar{l}k}-
g^{k\bar{p}}\p_{\bar{p}}d^{\Phi_0}_{\bar{l}}g^{\bar{l}i}=0\quad and \quad c.c.\\
&&\p_i\p_{\b k}\Phi_0=0,
\end{eqnarray}
which leads to the final system:
\begin{eqnarray}
&&2g^{r\bar{l}}\p_r\p_{\bar{l}}g^{i\bar{k}}-2\p_r g^{i\bar{p}}\p_{\bar{p}}g^{r\bar{k}}-
g^{i\bar{l}}\p_{\bar{l}}d^{\Phi_0}_sg^{s\bar{k}}-g^{r\bar{k}}\p_r d^{\Phi_0}_{\bar{j}}g^{\bar{j}i}
+\nonumber\\
&&\p_rg^{i\bar{k}}d^{\Phi_0}_{\bar{j}}g^{\bar{j}r}+
\p_{\bar{p}}g^{\bar{k}i}d^{\Phi_0}_n g^{n\bar{p}}=0\\
&&\p_{\bar{p}}d^{\Phi_0}_{\bar{l}}g^{\bar{l}k}=0 \quad \p_{p}d^{\Phi_0}_{l}g^{l\b k}=0, \quad 
\p_i\p_{\bar{k}}\Phi_0=0.
\end{eqnarray}

\section*{Appendix C: Proof of the Proposition 5.1.}
\addcontentsline{toc}{section}{Appendix C}
Here we will give the expression for the Einstein equations 
\begin{eqnarray}
R_{\mu\nu}+2\nabla_{\mu}\p_{\nu}\Phi=0
\end{eqnarray} 
with 
the metric and a dilaton
\begin{eqnarray}
&&G_{\mu\nu}=\eta_{\mu\nu}-th_{\mu\nu}(X)-t^2s_{\mu\nu}(X)+O(t^3),\nonumber\\ 
&&\Phi= \Phi_0+t\Phi_1(X)+
t^2 \Phi_2(X)+O(t^3),
\end{eqnarray} 
expanded to the second order of perturbation parameter $t$.
At the first order in $t$ we have:
\begin{eqnarray}\label{h}
1/2\Delta h_{\mu\nu}-1/2\p_{\mu}\p_{\rho}h^{\rho}_{\nu}-1/2\p_{\nu}\p_{\rho}h^{\rho}_{\mu}+
1/2\p_{\mu}\p_{\nu}h+2\p_{\mu}\p_{\nu}\Phi_1=0,
\end{eqnarray} 
where $h=\eta^{\rho\sigma}h_{\rho\sigma}$ and $\Delta=\p_{\mu}\p^{\mu}$. The indices are raised and lowered  
by means of the flat metric $\eta^{\rho\sigma}$. 
The next order gives:
\begin{eqnarray}\label{s}
&&1/2\Delta s_{\mu\nu}-1/2\p_{\nu}\p^{\beta}s_{\beta\mu}-1/2\p_{\mu}\p^{\beta}
s_{\beta\nu}+\p_{\nu}\p_{\mu}(1/2s+2\Phi_2+1/8 h^{\rho\sigma}h_{\rho\sigma})\nonumber\\
&&+1/2(\p^{\beta}h_{\beta\xi}-\p_{\xi} (1/2h+2\Phi_1))
\eta^{\xi\rho}(\p_{\rho}h_{\nu\mu}-\p_{\mu}h_{\nu\rho}-\p_{\nu}h_{\mu\rho})\nonumber\\
&&+1/2\eta^{\xi\rho}\p_{\xi}h_{\nu\lambda}\eta^{\lambda\alpha}\p_{\rho}h_{\alpha\mu}+
1/2\eta^{\xi\rho}\eta^{\sigma\alpha}h_{\rho\sigma}\p_{\xi}\p_{\alpha}h_{\mu\nu}\nonumber\\
&&-1/2\eta^{\xi\rho}\eta^{\lambda\alpha}\p_{\lambda}h_{\xi\nu}\p_{\rho}h_{\alpha\mu}-
1/2\p_{\sigma}\p_{\mu}h_{\nu\chi}h_{\xi\alpha}\eta^{\xi\chi}\eta^{\sigma\alpha}-\nonumber\\
&&1/2\p_{\sigma}\p_{\nu}h_{\mu\chi}h_{\xi\alpha}\eta^{\xi\chi}\eta^{\sigma\alpha}+
1/4h_{\alpha\rho}\p_{\nu}\p_{\mu}h_{\xi\lambda}\eta^{\alpha\xi}\eta^{\rho\lambda}=0.
\end{eqnarray}
Now we will obtain these equations from the master equation described above (we are interested in the 
terms of the  order $\alpha'^0$).
We have:
\begin{eqnarray}
&&V_1=1/2\alpha'^{-1}h_{\mu\nu}(X)\p X^{\mu}\bp X^{\nu},\\
&&V_2=1/2\alpha'^{-1}(s_{\mu\nu}(X)+
1/2h_{\mu\rho}\eta^{\rho\sigma}h_{\nu\sigma}(X))\p X^{\mu}\bp X^{\nu}.
\end{eqnarray} 
So, we just need to substitute these operators in the equations \rf{m1}, \rf{m2}. 
Starting from the first one
\begin{eqnarray}
&&(L_0V_1)-V_1-1/2L_{-1}L_1V_1-1/2\b L_{-1}\b L_1V_1-\nonumber\\
&&1/2L_{-1}\b L_{-1} (U_1+\b U_1)=0,
\end{eqnarray} 
we see that 
\begin{eqnarray}
&&(L_0V_1)-V_1=-1/4\Delta h_{\mu\nu}(X)\p X^{\mu}\bp X^{\nu},\\
&&1/2(L_1V_1+\b L_{-1}U_1)=-1/4(\p^{\beta}h_{\beta\xi}-2\p_{\xi}U_1)\bp X^{\xi},\\
&&1/2(\b L_1V_1+L_{-1} \b U_1)=-1/4(\p^{\beta}h_{\beta\xi}-2\p_{\xi}\b U_1)\p X^{\xi}.
\end{eqnarray} 
In such a way we see, that this equation coincides with \rf{h} if
\begin{eqnarray}
U_1+\b U_1=1/2h+2\Phi_1.
\end{eqnarray} 
The equation \rf{m2} is more complicated and should give \rf{s}:
\begin{eqnarray}\label{sa}
&&(L_0V_2)(z)-V_2(z)-1/2(V_1,V_1)_0^{(1,1)}(z)-1/2
( L_1V_1+\b L_{-1} U_1,V_1)_0^{(0,1)}(z)-\nonumber\\ 
&& 1/2(\b L_1V_1+L_{-1} \b U_1,V_1)_0^{(1,0)}(z)+L_{-1}\b W_2(z)+ \b L_{-1} W_2(z)=0,\nonumber\\
&&\b W_2(z)=-1/2((L_1 V_2)(z)-(L_1V_1+\b L_{-1} U_1,V_1)_0^{(1,1)}(z)-\nonumber\\
&&(\b L_1V_1+L_{-1} \b U_1,V_1)_0^{(2,0)}(z)+(U_1,V_1)_0^{(1,0)}(z)+\b L_{-1}U_2(z))\nonumber\\
&&W_2(z)=-1/2((\b L_1 V_2)(z)-(\b L_1V_1+L_{-1} \b U_1,V_1)_0^{(1,1)}(z)-\nonumber\\
&&(L_1V_1+\b L_{-1} U_1,V_1)_0^{(0,2)}(z)+(\b U_1,V_1)_0^{(0,1)}(z)+ L_{-1}\b U_2(z)).
\end{eqnarray} 
The $W$-terms are:
\begin{eqnarray}\label{first}
&&W_2=1/8\p^{\xi}(h_{\alpha\beta}\eta^{\beta\nu}h_{\nu\xi}+2s_{\xi\alpha})\p X^{\alpha}+\\ 
&&1/8(\p^{\beta}h_{\beta\xi}-2\p_{\xi}(U_1+\b U_1))
\eta^{\xi\rho}h_{\rho\alpha} \p X^{\alpha}-1/2\p_{\alpha}\b U_2\p X^{\alpha},\nonumber\\
&&\b W_2=1/8\p^{\xi}(h_{\alpha\beta}\eta^{\beta\nu}h_{\nu\xi}+2s_{\xi\alpha})\bp X^{\alpha}+\\
&&1/8(\p^{\beta}h_{\beta\xi}-2\p_{\xi}(U_1+\b U_1))\eta^{\xi\rho}h_{\rho\alpha} \bp X^{\alpha}-
1/2\p_{\alpha} U_2\bp X^{\alpha}\nonumber.
\end{eqnarray}
Here are the explicit formulas for other terms in sum \rf{sa}:
\begin{eqnarray}
&&(L_0-1)V_2=(L_0-1)(1/2\alpha'^{-1}s_{\mu\nu}\p X^{\mu}\bp 
X^{\nu})+\nonumber\\
&&(L_0-1)(1/4\alpha'^{-1}h_{\mu\beta}\eta^{\beta\alpha}h_{\nu\alpha}\p X^{\mu}\bp 
X^{\nu})\\
&&(L_0-1)(1/2\alpha'^{-1}s_{\mu\nu}\p X^{\mu}\bp 
X^{\nu})=-1/4\Delta s_{\mu\nu}\p X^{\mu}\bp X^{\nu}, \\
&&(L_0-1)(1/4\alpha'^{-1}h_{\mu\beta}\eta^{\beta\alpha}h_{\nu\alpha}\p X^{\mu}\bp 
X^{\nu})=\\
&&-1/8\Delta h_{\mu\nu}\eta^{\nu\alpha}h_{\alpha\beta}\partial X^{\mu}
\bar{\partial}X^{\beta}-
1/8h_{\mu\nu}\eta^{\nu\alpha}\Delta h_{\alpha\beta}\partial X^{\mu}\bar{\partial}X^{\beta}\nonumber\\
&&-1/4\eta^{\xi\rho}\partial_{\xi}h_{\mu\nu}\eta^{\nu\alpha}\p_{\rho}h_{\alpha\beta}
\p X^{\mu}\bp X^{\beta}=-1/4\eta^{\xi\rho}\partial_{\xi}h_{\mu\nu}\eta^{\nu\alpha}\p_{\rho}h_{\alpha\beta}
\p X^{\mu}\bp X^{\beta}\nonumber\\
&&-1/8(\p_{\nu}\p^{\xi}h_{\xi\mu}\eta^{\nu\alpha}h_{\alpha\beta}+
\p_{\nu}\p^{\xi}h_{\xi\beta}\eta^{\nu\alpha}h_{\alpha\mu})
\partial X^{\mu}\bp X^{\beta}-\nonumber\\
&&1/8\p_{\nu}((\p^{\beta}h_{\beta\xi}-2\p_{\xi}(U_1+\b U_1))
\eta^{\xi\rho}h_{\rho\alpha} )\bp X^{\nu} \p X^{\alpha}-\nonumber\\
&&1/8\p_{\nu}((\p^{\beta}h_{\beta\xi}-2\p_{\xi}(U_1+\b U_1))
\eta^{\xi\rho}h_{\rho\alpha} )\p X^{\nu} \bp X^{\alpha}+\nonumber\\
&&1/8(\p^{\beta}h_{\beta\xi}-2\p_{\xi}(U_1+\b U_1))
\eta^{\xi\rho}\p_{\nu}h_{\rho\alpha}\bp X^{\nu} \p X^{\alpha}+\nonumber\\
&&1/8(\p^{\beta}h_{\beta\xi}-2\p_{\xi}(U_1+\b U_1))
\eta^{\xi\rho}\p_{\alpha}h_{\rho\nu}\bp X^{\nu} \p X^{\alpha},\nonumber
\end{eqnarray}
\begin{eqnarray}
&&(V_1,V_1)_0^{(1,1)}=\\
&&(4\alpha'^2)^{-1}(h_{\rho\sigma}\p X^{\rho}\bp X^{\sigma}, h_{\lambda\mu}\p X^{\lambda}\bp X^{\mu})^{(1,1)}=\nonumber\\
&&1/2\eta^{\xi\rho}\eta^{\sigma\alpha}h_{\rho\sigma}\p_{\xi}\p_{\alpha}h_{\mu\nu}
\p X^{\mu}\bp X^{\nu}-
1/2\eta^{\xi\rho}\eta^{\lambda\alpha}\p_{\lambda}h_{\xi\nu}\p_{\rho}h_{\alpha\sigma}\p X^{\nu}\bp X^{\sigma}-
\nonumber\\
&&1/4\p_{\sigma}\p_{\rho}h_{\mu\nu}h_{\xi\alpha}\eta^{\xi\nu}\eta^{\sigma\alpha}\p X^{\mu}\bp X^{\rho}-
1/4\p_{\sigma}\p_{\rho}h_{\mu\nu}h_{\xi\alpha}\eta^{\xi\nu}\eta^{\sigma\alpha}\p X^{\rho}\bp X^{\mu}+\nonumber\\
&&1/4\p_{\rho}h_{\mu\nu}\p_{\sigma}h_{\xi\alpha}\eta^{\xi\nu}\eta^{\sigma\mu}\p X^{\rho}\bp X^{\alpha}+
1/4\p_{\rho}h_{\mu\nu}\p_{\sigma}h_{\xi\alpha}\eta^{\xi\nu}\eta^{\sigma\mu}\p X^{\alpha}\bp X^{\rho}+\nonumber\\
&&1/4h_{\alpha\rho}\p_{\nu}\p_{\mu}h_{\xi\lambda}\eta^{\alpha\xi}\eta^{\rho\lambda}\p X^{\mu}\bp X^{\nu}+O(\alpha'),
\nonumber
\end{eqnarray}
\begin{eqnarray}
&&-1/2(\b L_1V_1+L_{-1}\b U_1,V_1)_0^{(1,0)}=\\
&&(8\alpha')^{-1}((\p^{\beta}h_{\beta\xi}-2\p_{\xi}\b U_1)\p X^{\xi}, h_{\mu\nu} \p X^{\mu}\bp X^{\nu})^{(1,1)}=\nonumber\\
&&-1/8\p_{\rho}(\p^{\beta}h_{\beta\xi}\eta^{\xi\alpha}h_{\alpha\beta})\p X^{\rho}\bp X^{\beta}+
1/8\p^{\beta}h_{\beta\xi}\eta^{\xi\alpha}\p_{\rho}h_{\alpha\beta}\p X^{\rho}\bp X^{\beta}+\nonumber\\
&&1/8\p_{\lambda}\p^{\beta}h_{\beta\rho}\eta^{\lambda\alpha}h_{\alpha\beta}\p X^{\rho}\bp X^{\beta}
-1/8(\p^{\beta}h_{\beta\xi}-2\p_{\xi}\b U_1)\eta^{\xi\lambda}\p_{\lambda}h_{\rho\beta}\p X^{\rho}\bp X^{\beta}+\nonumber\\
&&O(\alpha'),
\nonumber
\end{eqnarray}
\begin{eqnarray}\label{last}
&&-1/2(L_1V_1+\b L_{-1}U_1,V_1)_0^{(0,1)}=\\
&&(8\alpha')^{-1}((\p^{\beta}h_{\beta\xi}-2\p_{\xi}\b U_1)\bp X^{\xi}, h_{\mu\nu} \p X^{\mu}\bp X^{\nu})^{(1,1)}=
\nonumber\\
&&-1/8\p_{\rho}(\p^{\beta}h_{\beta\xi}\eta^{\xi\alpha}h_{\alpha\beta})\bp X^{\rho}\p X^{\beta}+
1/8\p^{\beta}h_{\beta\xi}\eta^{\xi\alpha}\p_{\rho}h_{\alpha\beta}\bp X^{\rho}\p X^{\beta}+\nonumber\\
&&1/8\p_{\lambda}\p^{\beta}h_{\beta\rho}\eta^{\lambda\alpha}h_{\alpha\beta}\bp X^{\rho}\p X^{\beta}
-1/8(\p^{\beta}h_{\beta\xi}-2\p_{\xi}\b U_1)\eta^{\xi\lambda}\p_{\lambda}h_{\rho\beta}\bp X^{\rho}\p X^{\beta}+\nonumber\\
&&O(\alpha').\nonumber
\end{eqnarray} 
Collecting formulae \rf{first}-\rf{last} in \rf{sa} we arrive to the Einstein equations \rf{s}, putting 
\begin{eqnarray}
U_2+\b U_2=1/2s+2\Phi_2+1/8h_{\mu\nu}h^{\mu\nu}.
\end{eqnarray}

\end{document}